\documentclass[aps]{revtex4}
\usepackage{epsfig,graphicx,subfigure,amsmath,amssymb,bm}
\begin{document}

\newcommand{\be}{\begin{equation}}
\newcommand{\ee}{\end{equation}}
\newcommand{\bea}{\begin{eqnarray}}
\newcommand{\eea}{\end{eqnarray}}
\newcommand{\nn}{\nonumber}
\newcommand{\ba}{\bea \begin{array}}
\newcommand{\ea}{\end{array} \eea}
\renewcommand{\(}{\left(}
\renewcommand{\)}{\right)}
\renewcommand{\[}{\left[}
\renewcommand{\]}{\right]}
\newcommand{\bc}{\begin{center}}
\newcommand{\ec}{\end{center}}
\newcommand{\pa}{\partial}

\newcommand{\mb}[1]{ \mbox{\boldmath$#1$} }
\newcommand{\ds}{\displaystyle}
\newcommand{\beq}{\begin{eqnarray}}
\newcommand{\eeq}{\end{eqnarray}}
\newcommand{\beqq}{\begin{eqnarray*}}
\newcommand{\eeqq}{\end{eqnarray*}}
\newcommand{\p}{\partial}
\newcommand{\g}{\gamma}
\newcommand{\eps}{\varepsilon}
\newcommand{\x}{\mbox{\boldmath$x$}}
\newcommand{\n}{\mbox{\boldmath$n$}}
\newcommand{\J}{\mbox{\boldmath$J$}}
\newcommand{\y}{\mbox{\boldmath$y$}}
\newcommand{\w}{\mbox{\boldmath$w$}}
\newcommand{\z}{\mbox{\boldmath$z$}}

\title{Estimating the rate constant of cyclic GMP hydrolysis by activated phosphodiesterase in photoreceptors}

\author{J\"urgen Reingruber }
\affiliation{Department of Computational Biology, Ecole Normale Sup\'erieure, 46 rue
d'Ulm 75005 Paris, France.}
\author{David Holcman}
%\email[e-mail:]{holcman@biologie.ens.fr}
\affiliation{Department of Applied Mathematics, Weizmann Institute of
Science, Rehovot 76100, Israel}
\affiliation {Department of Computational Biology, Ecole Normale Sup\'erieure, 46 rue d'Ulm
75005 Paris, France}

\begin{abstract}
The early steps of light response occur in the outer segment of rod and cone photoreceptor. They
involve the hydrolysis of cGMP, a soluble cyclic nucleotide, that gates ionic channels located in
the outer segment membrane. We shall study here the rate by which cGMP is hydrolyzed by activated
phosphodiesterase (PDE). This process has been characterized experimentally by two different rate
constants $\beta_d$ and $\beta_{sub}$: $\beta_d$ accounts for the effect of all spontaneously
active PDE in the outer segment, and $\beta_{sub}$ characterizes cGMP hydrolysis induced by a
single light-activated PDE. So far, no attempt has been made to derive the experimental values of
$\beta_d$ and $\beta_{sub}$ from a theoretical model, which is the goal of this work. Using a model
of diffusion in the confined rod geometry, we derive analytical expressions for $\beta_d$ and
$\beta_{sub}$ by calculating the flux of cGMP molecules to an activated PDE site. We obtain the
dependency of these rate constants as a function of the outer segment geometry, the PDE activation
and deactivation rates and the aqueous cGMP diffusion constant. Our formulas show good agreement
with experimental measurements. Finally, we use our derivation to model the time course of the cGMP
concentration in a transversally well stirred outer segment.
\end{abstract}

\maketitle

%
%%%%%%%%%%%%%%%%%%%%%%%%%%%%%%%%%%%%%%%%%%%%%%%%%%%%%%%%%%%%%%%
\section{Introduction}
%%%%%%%%%%%%%%%%%%%%%%%%%%%%%%%%%%%%%%%%%%%%%%%%%%%%%%%%%%%%%%%
%
The modern theory of chemical reactions originates back to Arrhenius \cite{Arrhenius1889}, who
showed in 1889 that the backward rate constant $k_b$ of two reactants depends exponentially on the
temperature and the activation energy barrier. However, the molecular description of the backward
rate started with the seminal paper of Kramers in 1940
\cite{Kramers1940} (see also \cite{Haenggietal1990,BookSchuss}). The constant $k_b$ is used to
describe the chemical reaction of abundant species in solution and the concentration  of the
product resulting of the interaction of two molecules is calculated by using the mass action law.
But, The computation also involves the forward ${k_f}$ rate and the concentration of the two
species. At a molecular level, ${k_f}$ reflects the mean time for one of molecule to meet the other
by diffusion, and the probability to react upon encounter. { For diffusion limited chemical
reactions, based on the mean time for a uniform concentration of particles inside an infinite
3-dimensional space to hit a sphere of radius $a$, von Smoluchowski obtained in 1914 the first
estimate ${k_f}=4 \pi a D$
\cite{Smoluchowski1914}.} The Smoluchowski formula
was later on extended to the case of a partially absorbing sphere
\cite{BergPurcell1977,ZwanzigPNAS1990} .

{Diffusion plays in many cases a prominent role in the determination of the forward binding rate
\cite{SzaboSchultenSchulten1980,SzaboSchultenSchulten1981,PericoBattezzati1981,WilemskiFixman1973,CollinsKimball1949}},
and numerous fundamental processes in cellular biology rely on the rate at which diffusing
molecules hit a small target site: examples are trapping in patchy surfaces
\cite{Berezhkovskiietal2004}, receptor dwell time inside a synapse
\cite{TafliaHolcman2007} and many more. When the number of molecules is not large, the mass action law is
not sufficient to account for the random nature of the chemical
reactions and other approaches are required
\cite{HolcmanSchuss2005_JCP}. In addition, in a confined geometry,
the Smoluchowski formula does not describe the refine structure of
the bounded space. For that purpose, the small hole approximation
was developed, which is the mean time for a Brownian particle to
escape a confined domain through a small window
\cite{Grigorievetal2002,HolcmanSchuss2005_JCP,HolcmanPNAS2007}.
However, all these computations rely on the assumption that the
reaction volume is quite homogenous and has a shape close to a
convex domain (no bottle neck).

In photoreceptor outer-segment, a diffusing molecule needs to find
a specific target site in a degenerated domain, where one
dimensional length is much smaller than the others, and thus
previous related to the small hole formula do not apply. This
problem contains two difficulties: first, the target site occupies
only a tiny portion of the boundary, and second, the diffusion
occurs in a narrow domain.

We shall now be specific and explain what is our goal in the context of phototransduction: Rod
photoreceptors are highly specialized biological devices that can detect a single photon absorption
\cite{Hechtetal1942,Sakitt1972,BaylorLambYau1979,RevRiekeBaylor1998}.
The photon absorption activates a cascade of chemical reactions in
the outer segment, which ultimately hyperpolarizes the cell
\cite{RevPughLamb1993,RevRiekeBaylor1998,RevPughLamb2000,RevBurnsBaylor2001,RevArshavskyLambPugh2002,RevBurnsArshavsky2005}.
The inner structure of a rod outer segment is very specific and can be considered as a cylinder
that contains a densely packed stack of parallel and uniformly distributed discs (see
Fig.~\ref{Rod_OS_geometry}). {The discs divide the outer segment into almost separate compartments
that are loosely connected through a narrow gap between the disc perimeters and the outer segment
membrane, which we refer to as the outer shell
\cite{DiBenedetto2003}. Compartments are also linked through disc
incisures, however, since their impact is small \cite{DiBenedetto2006} we will neglect them in
first approximation }. The chemical reactions involved in the early steps of phototransduction
occur on the surface of the internal discs, and result in the activation of the phosphodiesterase
molecule (PDE) via a G-protein coupled activation cascade
\cite{RevPughLamb1993,RevRiekeBaylor1998,RevPughLamb2000,RevBurnsArshavsky2005}.
A photon-excited rhodopsin activates many transducin molecules,
which bind to and thereby activate PDE. The number of activated
PDE molecules following a single photon absorption was studied
both experimentally and theoretically
\cite{PughLamb1992,RiekeBaylor1996,Felber1996,Leskovetal2000,Hamer2003,ReingruberHolcman2007}. We
refer to PDE molecules that become activated via the
phototransduction cascade as light-activated PDE. In addition to
the transduction pathway, PDE can also spontaneously activate,
leading to a non-vanishing background activity even in darkness
\cite{RiekeBaylor1996,HolcmanKorenbrot2005}.

Cytoplasmic diffusible cGMP molecules controlling the opening of
cationic channels in the plasma membrane are hydrolyzed by
activated PDE, and the reduction in the cGMP concentration leads
to channel closure and photoreceptor hyperpolarization. From
another chemical pathway catalyzed by guanyl cyclase (GC), a
molecule attached to the disc surfaces and the outer segment
membrane, cGMP molecules are synthesized from cytoplasmic GTP, a
reaction which is calcium dependent. The magnitude of the
photoresponse signal depends significantly on the number of closed
ionic channels, and therefore on the drop in the cGMP
concentration, which is controlled in part by the number of
activated PDE and the rate of GMP hydrolysis of a single activated
PDE.

cGMP hydrolysis is characterized by two rate constants $\beta_d$ and $\beta_{sub}$, which are both
derived from experimental measurements
\cite{Hamer2005,Hamer2003,DiBenedetto2005,DiBenedetto2003,RevPughLamb2000,RiekeBaylor1996,RevPughLamb1993}.
Our goal here is to derive these constants from molecular considerations and biophysical theory,
and thus obtain explicit analytical expressions. To understand at an intuitive level how these
rates are defined, we recall that in most photoresponse models the cGMP concentration in the outer
segment is well-stirred, a simplification that neglects diffusion and the complex geometry of the
outer segment. The effective differential equation for the well-stirred cGMP concentration $C(t)$
is
\cite{RevPughLamb1993,RevPughLamb2000,Hamer2003}
\beq \label{empEqcGMP}
\frac{d}{dt} C(t) = \alpha(t) - \beta_d C(t) -\beta_{sub} P_l^*(t) C(t)\,,
\eeq
where $P_l^*(t)$ is the number of light-activated PDE molecules and $\alpha(t)$ the rate of cGMP
synthesis. The term $\beta_d C(t)$ accounts for cGMP hydrolysis due to spontaneous PDE activation,
and $\beta_{sub} P_l^*(t) C(t)$ due to light-activated PDE. Eq.~\ref{empEqcGMP} shows an important
difference in modeling cGMP hydrolysis by spontaneously- and light-activated PDE: whereas $\beta_d$
is the rate constant for the change in the cGMP concentration due to all spontaneously activated
PDE in the outer segment, $\beta_{sub}$ denotes the change in the well stirred cGMP concentration
due to a single light-activated PDE.

In the literature, $\beta_d$ and $\beta_{sub}$ are considered as independent parameters, a
hypothesis that is strengthened by the finding that the experimental values for $\beta_d$ and
$\beta_{sub}$ are around $1s^{-1}$ and $10^{-4}s^{-1}$ respectively, and therefore are extremely
different in appearance
\cite{RevPughLamb2000,RevPughLamb1993,Leskovetal2000,RevArshavskyLambPugh2002}.

{Based on the diffusional encounter process between a cGMP and an activated PDE molecule in the
complex rod outer segment geometry, we obtain explicit estimates for $\beta_d$ and $\beta_{sub}$.
Our analysis is motivated by several known results: First, in darkness, in average around one
spontaneously activated PDE molecule is present in a single compartment
\cite{RiekeBaylor1996,RevPughLamb2000}, which suggests that diffusion is rate limiting for
hydrolysis. Second, experimental data
\cite{Leskovetal2000,RevArshavskyLambPugh2002} indicate that activated PDE is a nearly perfect
effector enzyme and hydrolyzes cGMP with a very high efficiency, which also hints that
cGMP-hydrolysis is limited by diffusion. Third, a diffusion limited hydrolysis reaction couples the
cytosolic cGMP level most strongly to the activation status of PDE, which is at the basis of
photoreceptor adaptation \cite{RevFainMatthews2001,RevPughLamb2000}.

One of the main results of this paper is formula
\ref{Eqforbetad2},
\bea
\beta_d =  D_{cG}  \frac{2 \pi \rho \mu_+}{\mu_-}\frac{8}{4\ln(\frac{R}{a})-3}\,,
\eea
which relates $\beta_d$ to the spontaneous PDE activation and deactivation rates $\mu_+$ and
$\mu_-$, the PDE surface density $\rho$, the effective reaction radius $a$, the radius $R$ of the
outer segment, and the cytoplasmic cGMP diffusion constant $D_{cG}$. Furthermore, by comparing this
purely diffusional cGMP hydrolysis rate to experimental results, we can estimate the impact of the
details of the chemical hydrolysis reaction.}

The paper is organized as follows: we first determine the rate constant of cGMP hydrolysis due to a
single activated PDE as a function of the cGMP concentration, the cGMP diffusion constant and the
geometrical structure of the outer segment. Using this result, we then compute the analytical
expressions for $\beta_d$ and $\beta_{sub}$. We find that $\beta_d$ is proportional to the mean
number of spontaneously active PDE molecules in a compartment and not in the whole outer segment.
We compare our analytical estimations with experimental measurements, and find good agreement. Our
analysis suggests that the main reason for the discrepancy between $\beta_d$ and $\beta_{sub}$ is
their incompatible definitions. By deriving $\beta_d$ and $\beta_{sub}$ from molecular events, we
show that they are no longer two independent parameters. Finally, we use our analysis to model the
spatio-temporal time course of a photoresponse in a transversally well-stirred outer segment.

%
%%%%%%%%%%%%%%%%%%%%%%%%%%%%%%%%%%%%%%%%%%%%%%%%%%%%%%%%%%%
\section{Rate of cGMP hydrolysis by activated PDE}
%%%%%%%%%%%%%%%%%%%%%%%%%%%%%%%%%%%%%%%%%%%%%%%%%%%%%%%%%%%
%

In this section, we estimate cGMP hydrolysis rate constant by a driven by single activated PDE
molecule $P^*$ when diffusion is the limiting step.  Later on, we use this result to derive
expressions for $\beta_d$ and $\beta_{sub}$. To illustrate our approach, we start with the
molecular model for cGMP hydrolysis:
\bea\label{chemreact1}
cGMP + P^*    \quad
\underset{k_{b}}{\overset{k_f}{\rightleftarrows}} \quad cGMP\!\cdot\! P^*
\quad
\overset{k_2}{\longrightarrow} \quad P^* + GMP\,.
\eea
{A cGMP molecule binds to a $P^*$ molecule with a forward rate $k_f$ and forms an intermediate
complex $cGMP\cdot P^*$. This complex can either dissociate with a backward rate $k_{b}$, or cGMP
becomes hydrolyzed to GMP with a rate $k_2$. We are interested in the rate $k_h$ by which cGMP
molecules are hydrolyzed, which, at steady state, balances the $cGMP$ production rate. In the
restricted rod outer segment, the overall forward binding rate is $k_f G_c$, where $G_c$ is the
number of cGMP molecules in a single compartment. As an example, in darkness, $G_c$ is roughly in
the range 100-1000, depending on the radius of the outer segment \cite{RevPughLamb2000}. From
Eq.~\ref{chemreact1}, using the overall forward binding rate and Michaelis-Menton approximation, we
obtain
\bea\label{eq_MM}
k_h= \frac{k_2 k_f G_c P^*}{k_2 +k_b +k_f G_c}\,.
\eea
In the physiological range of cGMP concentrations, we assume that $k_2\gg k_f G_c$, which implies
that the hydrolysis of $cGMP\cdot P^*$ proceeds much faster compared to the formation of a new
complex. Furthermore, since $P^*$ hydrolyzes cGMP with very high efficiency
\cite{Leskovetal2000,RevArshavskyLambPugh2002}), we suppose that $k_2\gg k_{b}$, and therefore
neglect the backward rate. Under these circumstances, Eq.~\ref{eq_MM} reduces to
\bea\label{chemreact2}
k_h = k_f G_c P^* \,,
\eea
which has exactly the form of the hydrolysis term in Eq.~\ref{empEqcGMP}. Eq.~\ref{chemreact2} can
be formally obtained by setting $k_2=\infty$, which means that cGMP hydrolysis occurs
instantaneously after the formation of the the complex $cGMP\cdot P^*$. In contrast, if we assume
that $k_2$ is small ($k_2\ll k_f G_c$ and $k_b\ll k_f G_c$), using Eq.~\ref{eq_MM}, this implies
that hydrolysis proceeds independently of the cGMP concentration with a rate $k_2 P^*$, a scenario
that is not experimentally supported \cite{RevPughLamb2000}.

For large values $k_2$, the cGMP hydrolysis rate in Eq.~\ref{chemreact2} is determined by the
forward binding rate $k_f$, whose value depends on two parameters: the encounter rate $k_e$ of cGMP
molecules with the $P^*$ site, and the probability $p$ that $cGMP\cdot P^*$ is formed upon
encounter. The probability $p$ depends on (largely unknown) molecular properties of cGMP and
activated PDE. In order to extract the impact of diffusion on cGMP hydrolysis, we set $p=1$ and
presume that the complex $cGMP\cdot P^*$ is formed each time a cGMP molecule encounters $P^*$.
Finally, by neglecting the molecular details of activated PDE, we do not distinguish between
spontaneously- and light-activated PDE, and we consider only activated versus non-activated PDE. If
mainly diffusional issues are relevant for cGMP hydrolysis, then the catalytic activities of
spontaneously- and light-activated PDE should be very similar, as was already suggested by
experimental findings
\cite{RiekeBaylor1996}. }

\begin{figure}[ht!]
\vspace{0cm}
  \begin{center}
    \mbox{ \includegraphics[scale=0.4]{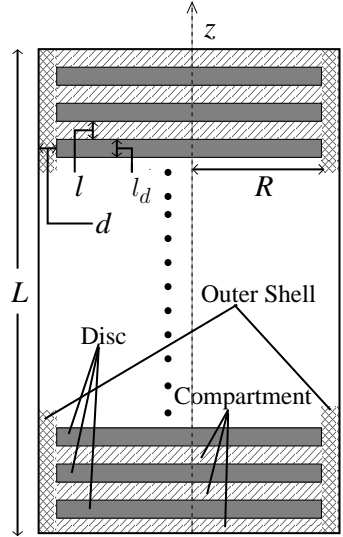} }
    \vspace{0cm}
    \caption{
    Section through a cylindrical rod outer segment,
    containing a densely packed stack of parallel and uniformly distributed discs.
    The volume delimited by two adjacent discs is called a compartment. }
    \label{Rod_OS_geometry}
  \end{center}
\end{figure}

Before starting the analysis, we give range values for the main
parameters: cGMP diffuses in the cytosolic volume $V_{cyto}$ of
the outer segment with a diffusion coefficient $D_{cG}\approx
100\mu m^2/s^{-1}$
\cite{Koutalosetal1995,OlsonPugh1993,DiBenedetto2005,HolcmanKorenbrot2005}.
In contrast, PDE molecules are attached to the disc surfaces,
where they diffuse with a diffusion coefficient $D_{PDE} \approx
0.8 \mu m^2/s^{-1}$ \cite{PughLamb1992}. The exact geometrical
dimensions of a rod outer segment varies between species
\cite{RevPughLamb2000,Nickeletal2007}: for example, the length $L$
and radius $R+d$ of the outer segment in a toad rod are $60\mu m$
and $3\mu m$, whereas in a mouse rod they are $20 \mu m$ resp.
$3\mu m$ \cite{RevPughLamb2000}. The longitudinal distance $l$
between two adjacent discs (the height of a compartment) and the
width of a disc $l_d$ (see Fig.~\ref{Rod_OS_geometry}) vary around
$15 nm$ \cite{Nickeletal2007}. The width $d$ of the outer shell is
comparable to $l$ \cite{Nickeletal2007}. The total number of
compartments $N_c=\frac{L}{l+l_d}$ in the outer segment is of the
order $N_c \sim 10^3$. Finally, we assume that the radii $a_{P}$
of a PDE molecule and $a_{cG}$ of a cGMP molecule are both
comparable to the radius of a rhodopsin molecule, which is around
1-2$nm$ \cite{PughLamb1992}. The parameters are summarized in
Table \ref{table_description}.

%%%%%%%%%%%%%%%%%%%%%%%%%%%%%%%%%%%%%%%%%%%%%%%%%%%%%%%%%%%%
\begin{table}
\small
\begin{center}
\begin{tabular}{|l|l|}
\hline
Symbol & Description  \\ [0.5ex] \hline
$L$ &  Length of a rod outer segment \\
$R$ &  Radius of a disc \\
$d$ &  Gap between disc and outer segment membrane \\
$l$ &  Distance between two adjacent disc \\
$l_d$ &  Width of a disc \\
$a_{P}$ &  Radius of a PDE molecule\\
$a_{cG}$ &  Radius of a cGMP molecule\\
$a=a_{P}+a_{cG}$ &  Sum of the radii of a cGMP and PDE molecule\\
$\rho$ & PDE surface density \\
$\mu_+$ & Spontaneous PDE activation rate \\
$\mu_-$ & Spontaneous PDE deactivation rate \\
\hline
\end{tabular}
\end{center}
\caption{Description of the parameters used in the model.}
\label{table_description}
\end{table}

%
%%%%%%%%%%%%%%%%%%%%%%%%%%%%%%%%%%%%%%%%%%%%%%%%%%%%%%%%%%%
\subsection{Analysis of cGMP hydrolysis due to a single activated PDE}
%%%%%%%%%%%%%%%%%%%%%%%%%%%%%%%%%%%%%%%%%%%%%%%%%%%%%%%%%%%
To describe the time course of cGMP concentration in the outer
segment, we consider different players: cGMP molecules are
independent and diffuse freely inside the outer segment domain
$\Omega$. Whenever a cGMP molecule hits the boundary area
$\partial \Omega_{h}$ occupied by the $P^*$ molecules, it becomes
instantaneously hydrolyzed. The synthesis of cGMP occurs on the
surface $\partial \Omega-\partial \Omega_{h}$ with a rate
$\alpha_\sigma(\x,t)$. We account for these interactions by using
the density $C(\x,t)$ of cGMP molecules at position $\x$ and time
$t$, it satisfies the diffusion equation with the appropriate
boundary condition \cite{BookSchuss},
\bea\label{diffEqgen}
\frac{\partial }{\partial t } C(\x,t)&=& D_{cG} \triangle C(\x,t)\,, \quad \mbox{for } \x \in
\Omega \\
D_{cG}\frac{\partial}{\partial n}C(\x,t) &=& -
\alpha_\sigma(\x,t) \,, \quad \mbox{  for } \x \in
\partial \Omega - \partial \Omega_{h} \label{boundCond1}\\
C(\x,t) &=& 0 \,, \quad   \mbox{for } \x \in
\partial \Omega_{h}\,.\label{boundCond2}
\eea
We shall now study Eq.~\ref{diffEqgen} for a single compartment $\Omega_c$.

%
%%%%%%%%%%%%%%%%%%%%%%%%%%%%%%%%%%%%%%%%%%%%%%%%%%%%%%%%%%%
\subsection*{Approximation of cGMP hydrolysis in a single compartment}
%%%%%%%%%%%%%%%%%%%%%%%%%%%%%%%%%%%%%%%%%%%%%%%%%%%%%%%%%%%
\label{AnalysiscGMPhydrolysis}
\begin{figure}[ht!] \vspace{0cm}
  \begin{center}
    \mbox{ \includegraphics[scale=0.9]{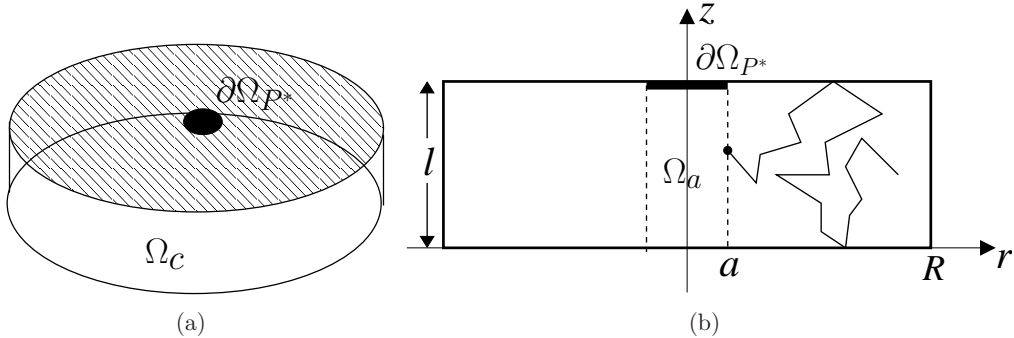} }
    \vspace{0cm}
    \caption{%\doublespacing
    (a) Elementary cylindrical compartment with a $P^*$ molecule located centrally on the upper surface.
    (b) Cross-section view of the compartment. A cGMP molecule is hydrolyzed when reaching the boundary of the volume $\Omega_a$ at $r=a$.}
    \label{fig_app_comp}
  \end{center}
\end{figure}
We now consider a compartment $\Omega_c$ in which a single PDE molecule is activated on either one
of the two disc surfaces (in Fig~\ref{FPT}a $P^*$ is attached to the upper surface). In our
approximation, cGMP hydrolysis rate is given by the flux $J_h$ of cGMP to the surface area $\p
\Omega_{P^*}$ occupied by a $P^*$ molecule. To compute $J_h$,  we shall make some approximations:

We consider a uniform and time independent cGMP synthesis rate
$\alpha_\sigma$. Because cGMP synthesis is calcium dependent, this
 corresponds to a calcium clamped outer segment or at equilibrium ( this
 is the case in darkness). Furthermore, because the height $l$ of a compartment is around a
few $nm$, and much smaller compared to the radius $R \sim
\mu m$, the time scale for longitudinal equilibration $l^2/D_{cG}$
is much shorter than the one for radial equilibration $\sim
R^2/D_{cG}$. Hence, newly synthesized cGMP molecules at the
surface quickly equilibrate in longitudinal direction before
encountering a $P^*$ molecule, which is usually located far away
compared to the compartment height $l$ (except for the negligible
amount cGMP synthesized in direct neighborhood of $P^*$). This
scenario is equivalent to having cGMP synthesized inside the
compartment, and we therefore replace cGMP synthesis on the
surface by synthesis inside the volume. The volume synthesis rate
$\alpha_v$ is linked to $\alpha_\sigma$ by
\bea
\alpha_v= 2\frac{\alpha_\sigma}{l},
\eea
where the factor 2 accounts for the two disc surfaces enclosing a
compartment. With a volume synthesis rate $\alpha_v$, the
diffusion equation for cGMP is
\bea\label{diffEqgen2}
\frac{\partial }{\partial t } C(\x,t)= D_{cG} \triangle C(\x,t) + \alpha_v \,, \quad \mbox{for } \x \in \Omega_c.
\eea
The boundary conditions are given by  Eq.~\ref{boundCond2} and Eq.~\ref{boundCond1} with
$\alpha_\sigma(\x,t)=0$.

{Because cGMP diffuses much faster than PDE ($D_{PDE}\ll D_{cG}$) we neglect PDE motion
\cite{PughLamb1992,Naqvi1974,TorneyMcConnell1983}. Since at leading order approximation the exact
position of the activated PDE is not relevant \cite{HolcmanetalNE1,HolcmanetalNE2,HolcmanetalNE3},
we position $P^*$ at the center of the disc.} We will discuss this issue in more detail in section
\ref{compExpRes}. In addition, we approximate cGMP molecules by infinitesimal points, and use the
effective reaction radius $a=a_{P}+a_{cG}$ for a $P^*$ molecule
\cite{PughLamb1992,Naqvi1974,TorneyMcConnell1983}.

Because the effective diameter $2a\sim 6nm$ of the boundary area $\partial \Omega_{P^*}$ occupied
by a $P^*$ molecule is comparable to the compartment height $l\sim 15nm$, and the radius $R\sim 3
\mu m$ is much larger than $a$ and $l$, the main limiting factor for cGMP hydrolysis rate is the
speed by which cGMP molecules find $P^*$. We note that once a cGMP molecule enters into a
neighborhood of $\partial \Omega_{P^*}$, since $2a$ is comparable to $l$, it has a high probability
to hit $\partial
\Omega_{P^*}$ and become hydrolyzed. In a first approximation, we
model the hydrolysis reaction by assuming that a cGMP molecule
entering the small cylindrical volume $\Omega_a$ (given in
cylindrical coordinates by $r\le a$) above or below
$\partial\Omega_{P^*}$ is instantaneously hydrolyzed (see
Fig.~\ref{fig_app_comp}). The corresponding boundary condition is
\bea \label{boundCond3}
C(\x,t) = 0\,, \quad \mbox{for } r=a\,.
\eea
This condition leads to an overestimation of the true hydrolysis
rate because cGMP molecules entering the domain $\Omega_a$ can as
well leave this region without touching the surface
$\partial\Omega_{P^*}$. However, in appendix
\ref{meanHydrolysisTime}, we show that the overestimation is in
the range of a factor 2 (see also the discussion in section
\ref{compExpRes}).

Having discussed the approximations, we shall now proceed to
estimate the cGMP flux $J_h$ into $\partial \Omega_a$. We are
particularly interested in $J_h$ as a function of the cGMP
concentration inside the compartment. From there, we will extract
cGMP hydrolysis rate due to a single activated PDE. With this
result, we will then derive the expression for $\beta_d$ (see
Eq.~\ref{empEqcGMP}). Using the cylindrical symmetry,
Eq.~\ref{diffEqgen2} reduces to
\bea\label{diffEq} \frac{\partial }{\partial t } C(r,t)&=& D_{cG}
\frac{1}{r} \frac{\partial }{\partial r } r \frac{\partial
}{\partial r } C (r,t)
+ \alpha_v \,,\\
 C (r,t)&=&0 \quad \mbox{for } r=a\,.
\eea
Integrating Eq.~\ref{diffEq} over the compartment volume yields an equation for the time dependent
number of cGMP molecules $G_c(t)$ in $\Omega_c$,
 \bea \label{eqGc1}
 \frac{d}{dt} G_c(t) = -J_R(t) - J_h(t) +\alpha_v \pi R^2 l \(1-\frac{a^2}{R^2}\)\,,
\eea
where
\bea
G_{c}(t) &=& 2 \pi l \int_{a}^{R}  C (r,t) r dr\,, \label{defGc}\\
J_{h}(t) &=& 2\pi a l  D_{cG}\frac{\partial}{\partial r}C(r,t)\Big |_{r=a}\,, \\
J_{R}(t) &=& - 2\pi R l D_{cG} \frac{\partial}{\partial r}C(r,t)\Big
|_{r=R} \,. \label{boundCondJR}
\eea
The flux $J_{R}(t)$ is maintained by cGMP molecules that diffuse
between compartments. {To derive an expression for $J_h$, we
consider the steady state regime where the flux $J_R$ is given. The
steady state concentration $C(r)$ obtained from Eq.~\ref{diffEq} is
given by
\bea
C(r) &=&   \frac{ \alpha_v \pi R^2 }{2 \pi D_{cG}} \(
\ln\(\frac{r}{a}\) -\frac{r^2-a^2}{2R^2}\)   - \frac{ J_R }{2\pi D_{cG}l}\ln\(\frac{r}{a}\) \,.
\label{equConc}
\eea
To obtain the number of $G_c$ molecules inside a compartment, we
insert Eq.~\ref{equConc} into Eq.~\ref{defGc} and for $\frac{a}{R}
\ll 1$, we obtain :
\bea\label{eqforGc}
G_c = \alpha_v \pi R^2 l \frac{R^2}{ D_{cG}}\(\frac{1}{2} \ln \(\frac{R}{a}\) -\frac{3}{8} \) -
J_R\frac{R^2}{D_G} \( \frac{1}{2} \ln\(\frac{R}{a}\) -\frac{1}{4} \) \,.
\eea
We define the times $\tau_1$ and $\tau_2$ and the corresponding
rates $k_1$ and $k_2$ as
\bea
\tau_1&=& \frac{1}{k_1} =
\frac{R^2}{D_{cG}}\(\frac{1}{2} \ln \(\frac{R}{a}\) -\frac{3}{8} \) \,, \label{tau1}\\
\tau_2&=& \frac{1}{k_2} = \frac{R^2}{D_{cG}} \( \frac{1}{2} \ln\(\frac{R}{a}\) -\frac{1}{4} \) \,,\label{tau2}
\eea
we can rewrite expression \ref{eqforGc} as
\bea\label{eqforGc2}
G_c = \frac{\alpha_v \pi R^2 l} {k_1} - \frac{J_R}{k_2} \,.
\eea
At steady state, the value of $J_h$ is fixed by the balance of fluxes, and Eq.~\ref{eqGc1} gives
for $\frac{a}{R} \ll 1$
\bea\label{equJa}
J_h = -J_R + \alpha_v\pi R^2 l \,.
\eea
Using Eq.~\ref{eqforGc2} we can express $\alpha_v$ as a function $G_c$ and $J_R$,
\bea\label{eqalpahv}
\alpha_v \pi R^2 l = k_1 G_c + \frac{k_1}{k_2} J_R \,.
\eea
Finally, inserting Eq.~\ref{eqalpahv} into Eq.~\ref{equJa} yields
\bea\label{fluxJa}
J_h = k_1 G_c + \frac{k_1-k_2}{k_2} J_R\,.
\eea
Formula \ref{fluxJa} gives the steady state hydrolysis rate $J_h$ as a function of $G_c$ and $J_R$.
This result depends strongly on the diffusional and geometrical properties of the microdomain.
Whereas Eq.~\ref{equJa} gives a direct expression for $J_h$ as a function of the synthesis rate
$\alpha_v$, and does not involve diffusion,  Eq.~\ref{fluxJa} is related to $\alpha_v$ indirectly
via the value of $G_c$, and therefore involves diffusion.}

In appendix \ref{meanHydrolysisTime} we obtain an interpretation for the two times $\tau_1$ and
$\tau_2$, and, thus, for the rates $k_1$ and $k_2$: $\tau_1$ (see Eq.~\ref{meantime}) is the mean
time for uniformly distributed cGMP molecules to reach the absorbing boundary at $r=a$, given
reflecting boundary conditions at $r=R$; $\tau_2$ (see Eq.~\ref{meanTimeTauo(r)} for $r=R$) is the
mean time to reach $r=a$, when the initial position is uniformly distributed at $r=R$. In reality,
there is no reflecting boundary at $r=R$, however, a vanishing flux $J_R$ is mathematically
equivalent to a reflecting boundary condition at $r=R$.

%
%%%%%%%%%%%%%%%%%%%%%%%%%%%%%%%%%%%%%%%%%%%%%%%%%%%%%%%%%%%%%
\subsection{Derivation of the rate constant $\beta_d$ for spontaneous PDE activation}
%%%%%%%%%%%%%%%%%%%%%%%%%%%%%%%%%%%%%%%%%%%%%%%%%%%%%%%%%%%%%
%
{In darkness, spontaneous PDE activation leads to a uniform cGMP
hydrolysis in the outer segment
\cite{RevPughLamb2000,RevPughLamb1993,RiekeBaylor1996} with an
overall hydrolysis rate (see Eq.~\ref{empEqcGMP} integrated over the cytoplasmic volume)
\bea\label{overallJ}
J_{d,os}=\beta_d G_{os}\,,
\eea
where $G_{os}$ is the total number of cGMP molecules in the outer segment. To derive an analytical
expression for the rate constant $\beta_d$, we start from Eq.~\ref{fluxJa}.} Because spontaneous
PDE activation occurs uniformly throughout the outer segment, apart from fluctuations, the flux
$J_R$ between compartments vanishes in darkness. Thus, the steady state hydrolysis rate $J_h$ of a
single $P^*$ molecule given in Eq.~\ref{fluxJa} can be written as
\bea\label{fluxJa2}
J_h = k_1 G_c \,.
\eea
To obtain the dark hydrolysis rate $J_{d,c}$ per compartment, we
have to further consider the mean number of spontaneously
activated PDE molecules $P^*_{s,c}$ in a compartment. As long as
the number $P^*_{s,c}$ is small and the $P^*$ molecules are
geometrically well separated \cite{HolcmanSchuss2008_cluster}, the
rate $J_{d,c}$ increases linearly with $P^*_{s,c}$. Hence, we
obtain
\bea\label{eqJdc}
J_{d,c}=k_{1}P^*_{s,c} G_c\,.
\eea
The hydrolysis rate in the whole outer segment $J_{d,os}$ is obtained by summing $J_{d,c}$ over all
$N_c$ compartments. Since $G_{os}= N_c G_c$ well approximates the total number of cGMP molecules in
the outer segment (the volume $2\pi R d L$ of the outer shell is negligible compared to the volume
$\pi R^2 l N_c$ of all compartments), we obtain
\bea\label{eqJdos}
J_{d,os}= k_{1} P^*_{s,c} G_{os}\,.
\eea
Finally, by comparing Eq.~\ref{eqJdos} with Eq.~\ref{overallJ} and
by using Eq.~\ref{tau1}, we obtain
\bea\label{Eqforbetad}
\beta_d= k_{1} P^*_{s,c} =\frac{D_{cG}}{R^2}\frac{8}{4 \ln \(\frac{R}{a}\) -3}  P^*_{s,c}\,.
\eea
We conclude that $\beta_d$ is determined by the mean number of spontaneously activated PDE in a
compartment, and not in the  outer segment \cite{RiekeBaylor1996}. Furthermore, we will now relate
$\beta_d$ to the spontaneous PDE activation rate $\mu_+$, the deactivation rate $\mu_-$, and the
PDE surface density $\rho$. The number of PDE on the disc surfaces attached to a single compartment
is $P_c=2\pi R^2 \rho$, and $ P^*_{s,c}$ is given by
\bea\label{meanSpActPDE}
P^*_{s,c}= P_c \frac{\mu_+}{\mu_-} = 2 \pi R^2  \rho_{} \frac{\mu_+}{\mu_-} \,.
\eea
Together with Eq.~\ref{tau1} and Eq.~\ref{Eqforbetad} we obtain the final expression
\bea\label{Eqforbetad2}
\beta_d =  D_{cG}  \frac{2 \pi \rho \mu_+}{\mu_-}\frac{8}{4\ln(\frac{R}{a})-3}\,.
\eea

%
%%%%%%%%%%%%%%%%%%%%%%%%%%%%%%%%%%%%%%%%%%%%%%%%%%%%%%%%%%%%%
\subsection{Effective set of equations to model cGMP dynamics}
%%%%%%%%%%%%%%%%%%%%%%%%%%%%%%%%%%%%%%%%%%%%%%%%%%%%%%%%%%%%%
%
By generalizing our previous results, we shall now derive an effective set of equations to model
cGMP dynamics following a photon absorption. Since a photon absorbtion transiently generates an
elevated amount of $P^*$ molecules inside the affected compartment, it induces an increased cGMP
hydrolysis and a cGMP gradient in the outer segment. In this case, the fluxes $J_R$ between
compartments are no longer zero after a photon absorption.

We start the derivation by extending the equilibrium expression for $J_h$ given in Eq.~\ref{fluxJa}
to time dependent situations. Because free cGMP diffusion is fast, cGMP equilibrates quickly inside
a compartment. In contrast, $G_c(t)$ and $J_R(t)$ fluctuations are determined by the effective
longitudinal diffusion between the compartments, which is strongly hindered by the
compartmentalization of the outer segment
\cite{HolcmanKorenbrot2004,DiBenedetto2003,OlsonPugh1993}. Thus, we consider that $G_c(t)$ and
$J_R(t)$ fluctuate on a slower time scale compared to the equilibration time scale inside a
compartment. Under this condition, a first approximation of the time dependent hydrolysis rate
$J_h(t)$ of a single $P^*$ molecule is given by the equilibrium expression in Eq.~\ref{fluxJa} with
time dependent $J_R(t)$ and $G_c(t)$:
\bea
J_h(t) = k_{1} G_c(t) + \frac{k_1-k_2}{k_2} J_R(t)  \label{fluxJaoft}  \,.
\eea
{We note that the expression for $J_h$ given in Eq.~\ref{fluxJa} can be extended to time dependent
cases, whereas this is not possible starting from Eq.~\ref{equJa}.}

\begin{figure}[ht!] \vspace{1cm}
  \begin{center}
    \mbox{ \includegraphics[scale=0.8]{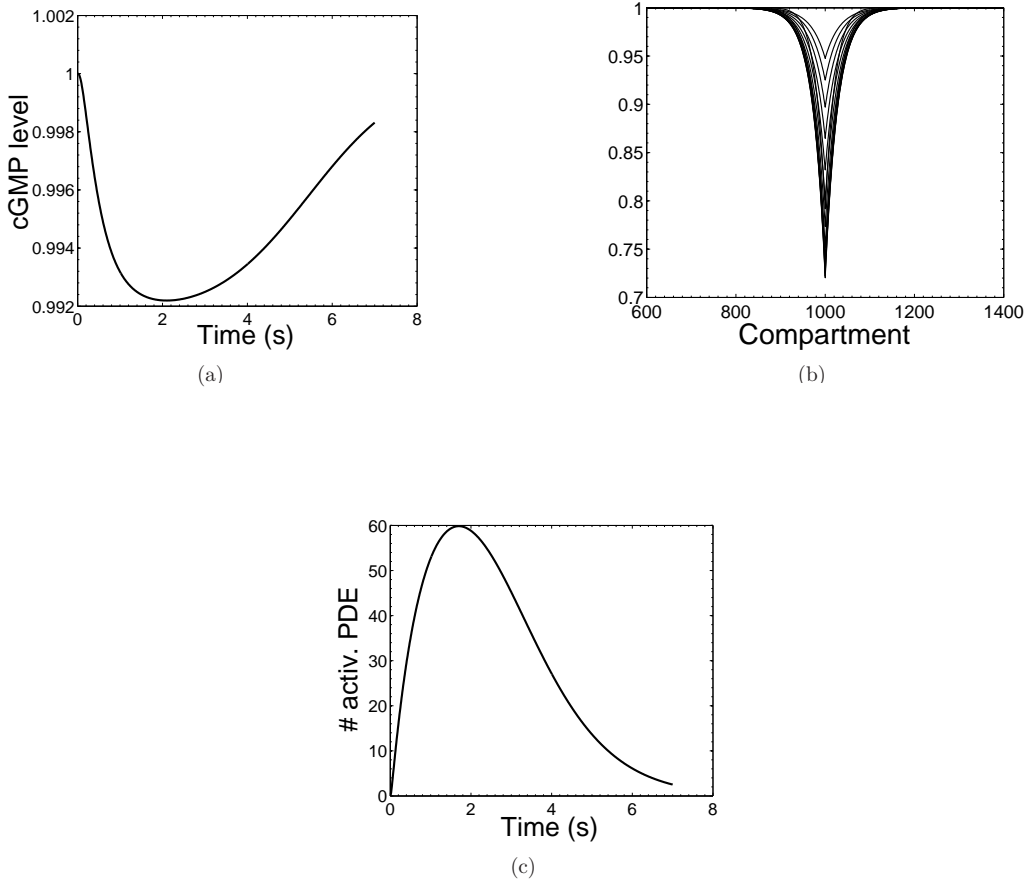} }
    \vspace{0cm}
    \caption{%\doublespacing
    cGMP dynamics after a photon absorption at time $t=0$ in compartment $n=1000$ ($N_c=2000$). The simulation is performed using Eqs.~\ref{GcwithPh} and \ref{J_R2}. The cGMP
concentration is scaled with the equilibrium value. (a) Time dependent cGMP level averaged over
 the outer segment. (b) cGMP level per compartment for various time points. (c) Number of
light-activated PDE molecules obtained using the equations published in
\cite{ReingruberHolcman2007}.}
    \label{fig_cGMP}
  \end{center}
\end{figure}

To obtain the set of equations for the time dependent number of cGMP molecules inside a
compartment, we start when a photon is absorbed in compartment $n_0$, while the other compartments
$n=1\ldots N_c$, $n\neq n_0$ remain unperturbed. In the regime considered here, cGMP hydrolysis
depends linearly on $P_l^*(t)$. Using Eq.~\ref{eqGc1}, the equation for the number $G_c^{(n)}(t)$
of cGMP molecules in a compartment $n$ is given by ($\delta_{n,n_0}$ is the Kronecker-Delta)
\bea
\frac{d}{dt} G_c^{(n)}(t) = -J_R^{(n)}(t) - ( P^*_{s,c} +
P_l^*(t)\delta_{n,n_0} ) J_h^{(n)}(t) + \alpha_v
\pi R^2 l\,. \label{GcwithPh0}
\eea
Inserting the expression for $J_h^{(n)}(t)$ given in Eq.~\ref{fluxJaoft}, and using the definition
of $\beta_d$ in Eq.~\ref{Eqforbetad}, we obtain
\bea
\frac{d}{dt} G_c^{(n)}(t) &=& -
\(1+( P^*_{s,c}  + P_l^*(t)\delta_{n,n_0} )\frac{k_1-k_2}{k_2}  \) J_R^{(n)}(t) \nn\\
&& - \beta_d G_c^{(n)}(t) - k_1 P_l^*(t) \delta_{n,n_0}  G_c^{(n)}(t)  + \alpha_v \pi R^2l \,.
\label{GcwithPh}
\eea
By approximating the transversally well stirred cGMP concentration in a compartment by
$C^{(n)}(t)\approx \frac{G_c^{(n)}(t)}{\pi R^2 l}$, and by using Fick's law, the fluxes
$J_R^{(n)}(t)$ are approximated by
\bea J_R^{(n)}(t)&=& - D_{cG}2\pi R d
\(\frac{C^{(n+1)}(t)-C^{(n)}(t)}{l+l_d} +
\frac{C^{(n-1)}(t)-C^{(n)}(t)}{l+l_d} \) \nn \\
&=& -D_{cG} \frac{2 d}{R l}
\(\frac{G_c^{(n+1)}(t)-G_c^{(n)}(t)}{l+l_d} +
\frac{G_c^{(n-1)}(t)-G_c^{(n)}(t)}{l+l_d} \)\,.\label{J_R2}
\eea
Eqs.~\ref{GcwithPh} and \ref{J_R2} constitute  a close system of equations for the $G_c^{(n)}(t)$
(which can be transformed into equations for the concentrations $C^{(n)}(t)$). Furthermore,
Eq.~\ref{GcwithPh}  models the impact of spontaneously- and light-activated PDE in an equivalent
way. The simulation in Fig.~\ref{fig_cGMP} shows the time course of the number of cGMP molecules,
scaled with respect to the dark equilibrium value, after the absorption of a photon at time $t=0$
in the middle of the outer segment. The parameters for the simulation are suitable for a toad rod
\cite{RevPughLamb2000}. The input function $P_l^*(t)$ is obtained using the set of equations
published in \cite{ReingruberHolcman2007}.

%
%%%%%%%%%%%%%%%%%%%%%%%%%%%%%%%%%%%%%%%%%%%%%%%%%%%%%%
\subsection{Derivation of the rate constant $\beta_{sub}$ in a well stirred outer segment}
%%%%%%%%%%%%%%%%%%%%%%%%%%%%%%%%%%%%%%%%%%%%%%%%%%%%%
%
We now derive an analytic expression for the rate constant $\beta_{sub}$ using the approximation of
a well stirred outer segment \cite{RevPughLamb2000}. Since the volume of the outer shell is
negligible compared to the combined volume of all compartments, the total number of cGMP molecules
in the outer segment is
\beq
G_{os}(t)= \sum_{n=1}^{N_c}G_c^{(n)}(t).
\eeq
By summing Eq.~\ref{GcwithPh} over all compartments, and using that $\sum_{n=1}^{N_c} J_R^{(n)}(t)
\approx 0$ (we neglect the cGMP molecules in the outer shell), we obtain
\bea\label{eqfortotG_0}
\frac{d}{dt} G_{os}(t) =
-\frac{k_1-k_2}{k_2} P_l^*(t) J_R^{(n_0)}(t) - \beta_d G_{os}(t) -
k_1 P_l^*(t) G_c^{(n_0)}(t) + \alpha_v \pi R^2 l N_c  \,.
\eea
In a well stirred outer segment we have $G_c^{(n)}(t) = G_{os}(t)/N_c$. By further neglecting the
term $-\frac{k_1-k_2}{k_2} P_l^*(t) J_R^{(n_0)}(t)$ (the flux $J_R$ vanishes in a well stirred
outer segment), we get
\bea\label{eqfortotG}
\frac{d}{dt} G_{os}(t) = - \beta_d G_{os}(t) -
\frac{k_1}{N_c} P_l^*(t) G_{os}(t) + \alpha_v \pi R^2 lN_c \,.
\eea
Finally, dividing Eq.~\ref{eqfortotG} with the cytosolic volume $V_{cyto}
\approx \pi R^2 l N_c$ yields the standard equation for the well stirred cGMP
concentration $C(t)= G_{os}(t)/V_{cyto}$,
\bea\label{eqforC}
\frac{d}{dt} C(t) = -  \beta_d  C(t)  - \frac{k_{1}}{N_c} P_l^*(t)
C(t) + \alpha_v \,.
\eea
By comparing Eq.~\ref{eqforC} with Eq.~\ref{empEqcGMP}  we obtain for $\beta_{sub}$ the expression
\bea\label{betasub}
\beta_{sub} =\frac{k_{1}}{N_{c}} =\frac{\beta_d}{N_{c} \bar
P^*_{s,c}}\,.
\eea
Since $N_c$ is of the order $10^3$, it follows that $\beta_{sub}$ is much smaller than $\beta_d$.
{ Using   Eq.~\ref{tau1} for $k_{1}$  and $V_{cyto}\approx \pi R^2 l N_c$, Eq.~\ref{betasub} can
be written as
\bea\label{betasub2}
\beta_{sub} = \frac{ \pi D_{cG} l}{V_{cyto}}\frac{8}{4 \ln \(\frac{R}{a}\)-3}\,.
\eea
By comparing expression \ref{betasub2} with the standard definition of $\beta_{sub}$ given by
\cite{RevPughLamb2000,RevPughLamb1993} (we neglect cytoplasmic buffering for cGMP
\cite{Leskovetal2000,DiBenedetto2005})
\bea\label{expdefbetasub}
\beta_{sub} = \frac{k_{sub}}{K_m N_{Av}V_{cyto}}\,,
\eea
we obtain a new formula for $\frac{k_{sub}}{K_m }$  given by ($N_{Av}$ is the Avogadro number)
\bea\label{ksubKm}
\frac{k_{sub}}{K_m } = \frac{N_{Av} V_{cyto} k_1}{N_c} = \frac{8\pi D _{cG} l N_{Av}}{4 \ln \(\frac{R}{a}\)-3} \,.
\eea
}
%
%%%%%%%%%%%%%%%%%%%%%%%%%%%%%%%%%%%%%%%%%%%%%%%%%%%%%%%
\section{Comparison with experimental results}
\label{compExpRes}
%%%%%%%%%%%%%%%%%%%%%%%%%%%%%%%%%%%%%%%%%%%%%%%%%%%%%%%%%
%
To validate our computations, we now compare our analytical results for $\beta_d$ and $\beta_{sub}$
(Eq.~\ref{Eqforbetad} and Eq.~\ref{betasub}) with experimental measurements
\cite{RevPughLamb2000,Hamer2003,DiBenedetto2003,RiekeBaylor1996}. We start with $\beta_d$.
Using data available for toad rods, $\rho=100 \mu m^{-1}$, $R=3\mu m$, $\mu_{+}=4 \times
10^{-4}s^{-1}$, $\mu_{-}=1.8 s^{-1}$, $D_{cG}=100\mu m^2/s^{-1}$, $a=3nm$, $l=15nm$
\cite{DiBenedetto2003,RevPughLamb2000,RiekeBaylor1996}, and inserting
these values into Eq.~\ref{tau1}, Eq.~\ref{meanSpActPDE} and Eq.~\ref{Eqforbetad}, we obtain
$k_{1}=3.6s^{-1}$, $P^*_{s,c}=1.26$ and
\bea
\beta_d = k_{1} P^*_{s,c} \approx 4.5 s^{-1}\,.
\eea
This analytic result has to be compared to the experimentally found value $\beta_d\approx 1s^{-1}$
\cite{RevPughLamb2000}, which is approximately four times smaller than this prediction.
Eq.~\ref{Eqforbetad2} shows that $\beta_d$ depends only logarithmically on the compartment radius
$R$, and thus it is very similar across species that differ mostly on the radius of the outer
segment, in agreement with experimental findings
\cite{RevPughLamb2000}. The discrepancy between our theoretical prediction and the experimental
value for $\beta_d$ can be attributed to several factors:
\begin{enumerate}
\item
We made the assumption that a cGMP molecule already becomes hydrolyzed when reaching the inner
cylinder $\Omega_a$ at $r=a$. Thus, the time $\tau_1$ in Eq.~\ref{tau1} is shorter than the true
time needed to arrive at the $P^*$ site. Hence, Eq.~\ref{tau1} overestimates the hydrolysis rate.{
In appendix \ref{meanHydrolysisTime}, we derive an accurate estimate for the mean time $\tau$ a
cGMP molecule reaches the $P^*$ site located on the surface of a compartment (see
Eq.~\ref{tauasymp}). Compared to $\tau_1$ (Eq.~\ref{tau1}), the new estimate for $\tau$ includes
specifically the mean time $\tau_a$ a cGMP molecule starting on the boundary of $\Omega_a$ reaches
the $P^*$ molecule on the surface. By considering the additional time $\tau_a$, we replace $\tau_1$
and $\tau_2$ with the more accurate expressions $\tilde \tau_1= \tau_a + \tau_1 $ and $\tilde
\tau_2= \tau_a + \tau_2$. Accordingly, the rates $k_1$ and $k_2$ have to be replaced by $\tilde
k_1$ and $\tilde k_2$, given by
\beq\label{rates_exact}
\tilde k_1= \frac{1}{\tilde \tau_1}  = \frac{1}{\tau_a +\tau_1} \,, \quad
\tilde k_2=  \frac{1}{\tilde \tau_2} =\frac{1}{\tau_a +\tau_2}\,.
\eeq
For toad rod values with $l/a \sim 5$ and $R/a \sim 1000$, and by using Eq.~\ref{tauasymp} with
$g(5)\approx 2.9$ (the value $g(5)$ is obtained from Fig.~\ref{fig_Taua}b), we find that $\tilde
k_1 \approx 0.5 k_1$. By using $\tilde k_1$ instead of $k_1$ in Eq.~\ref{Eqforbetad} we obtain the
new estimation
\bea\label{betadark_moreexact}
\beta_d=\tilde k_1 P^*_{s,c} = 2.25s^{-1} \,,
\eea
which is closer to the experimental observation.

\item
Our assumption that every encounter between cGMP and $P^*$ results in cGMP hydrolysis will
certainly lead to an overestimation of the hydrolysis rate. Moreover, since we neglected the
molecular details of the hydrolysis reaction, this will also induce an error. Nevertheless, since
our analytic result for $\beta_d$ is very close to the experimental finding, we conclude that cGMP
hydrolysis by $P^*$ has to be largely diffusion limited, and in addition has to be quite efficient,
such that nearly every encounter between cGMP and activated PDE leads to a hydrolysis reaction.
This is supported by the experimental observations that activated PDE hydrolyzes cGMP with very
high efficiency \cite{Leskovetal2000,RevArshavskyLambPugh2002}. }

\item
Uncertainties in the experimental values for $D_{cG}$, $\mu_+$ and $\mu_-$, involved in the
computation of $\beta_d$, introduce ambiguities in our analytical prediction. For example, there is
still considerable disagreements about the  exact value of the diffusion constant $D_{cG}$
\cite{Koutalosetal1995,OlsonPugh1993,DiBenedetto2005,HolcmanKorenbrot2005}. Furthermore, at first
approximation, we used for the effective reaction radius $a$ the sum of the molecular radii of a
PDE and cGMP molecule. A more precise value for $a$ will affect $\beta_d$ in
Eq.~\ref{betadark_moreexact} mainly via $\tau_a$, since $\tau_1$ depends only logarithmically on
$a$ (see Eq.~\ref{Eqforbetad2}.
%However, even if the effective reaction radius $a$, which is the sum of PDE and cGMP
%molecular radii is note precise, it will not affect much $\beta_d$ in Eq.~\ref{betadark_moreexact},
%since $\tau_1$ depends only logarithmically on $a$ (see Eq.~\ref{Eqforbetad2} and the contribution
%of $\tau_a$ is smaller.

\item
{The value of $\beta_d$ was computed by fixing the position of $P^*$ at the disk center and
neglecting possible fluxes between compartments. In general, spontaneous PDE activation and
diffusion leads to $P^*$ positions that are uniformly distributed over the disk surface, and
different $P^*$ positions in neighboring compartments induce small fluxes. We left open here the
computation of the variance of the cGMP hydrolysis rate constant coming from random locations of
$P*$ molecules. However, the $P^*$ position should not much influence the rate constant for cGMP
hydrolysis: The rate constant is determined by the MFPT of a cGMP molecule to find the $P^*$
target. Outside a small boundary layer around $P^*$ (the radius of the boundary layer is of the
order of the reaction radius $a$), the leading order term of the MFPT in dimension 2 depends only
logarithmically on the distance between cGMP and $P^*$, and in dimension 3 it is a constant
\cite{HolcmanetalNE1,HolcmanetalNE2}. Hence, since almost all cGMP molecules are outside the
boundary layer, the exact position of $P^*$ is not important for their mean time to hydrolysis. We
conclude that our expression for $\beta_d$ should remain a valid approximation at first order, even
when considering random $P^*$ positions.}

%\item
%{\bf The value of $\beta_d$ was computed by fixing the position of $P^*$ at the disk center and
%neglecting possible fluxes between compartments. In general, an arbitrary position will produce an
%variance in the value of $\beta_d$. The arbitrary position is generated by the spontaneous PDE
%activation and diffusion. But the steady state distribution of $P^*$ should be
% uniform  over the disk surface. We left open here the computation of cGMP hydrolysis rate
% constant variance coming from the random location of $P*$ molecules which should account for
%small fluxes between compartments generated by not centered $P^*$. However, the exact $P^*$
%position does not much influence the mean rate constant for cGMP hydrolysis: The rate constant is
%determined by the MFPT of a cGMP molecule to the $P^*$ target. Outside a small boundary layer
%around $P^*$ (the radius of the boundary layer is of the order of the reaction radius $a$), the
%leading order term of the MFPT in dimension 2 depends only logarithmically on the distance between
%cGMP and $P^*$, and in dimension 3 it is a constant \cite{HolcmanetalNE1,HolcmanetalNE2}. Hence,
%since almost all cGMP molecules are outside the boundary layer, the exact position of $P^*$ is not
%important for their mean time to hydrolysis. We conclude that our computation for $\beta_d$ should
%remain a valid approximation at first order even when considering random $P^*$ positions.}

\end{enumerate}

We shall now compare expressions Eqs.~\ref{betasub},\ref{betasub2} for $\beta_{sub}$, and the ratio
$\frac{k_{sub}}{K_m}$ (Eq.~\ref{ksubKm}) with experimental measurements. Eq.~\ref{betasub} reveals
that $\beta_{sub}$ is a factor $N_c P^*_{s,c}$ smaller than $\beta_d$. Since $N_c$ is of order
$10^3$ and $ P^*_{s,c}$ of order $1-10$, this agrees with the experimental findings that
$\beta_{sub}$ is around $10^3-10^4$ times smaller than $\beta_d$
\cite{RevPughLamb2000,RevArshavskyLambPugh2002}. From Eq.~\ref{ksubKm}, we obtain the
prediction
%(for the value $\frac{R}{a}$ we insert toad rod values)
\bea\label{ksubKmpred}
\frac{k_{sub}}{K_m} =  \frac{N_{Av} V_{cyto} k_1}{N_c} \approx 9.2 \times
10^8 M^{-1}s^{-1} \,. \nn
\eea
This estimation can be further improved by using the rate $\tilde k_1=0.5 k_1$ instead of $k_1$,
giving
\bea
\frac{k_{sub}}{K_m} \approx 4.6 \times 10^8 M^{-1}s^{-1}\,,\nn
\eea
which has to be compared to $\frac{k_{sub}}{K_m} \approx 2.2 \times 10^8 M^{-1}s^{-1}$ obtained
from experiment \cite{Leskovetal2000}. It is important to note that our analytic results for
$\frac{k_{sub}}{K_m}$ and $\beta_d$ (see Eq.~\ref{betadark_moreexact}) are both around two times
larger than the experimental findings, which is an indirect confirmation of our assumption that
$\beta_d$ and $\beta_{sub}$ (note that ${k_{sub}}/{K_m}$ is proportional to $\beta_{sub}$) are not
two independent rate constants, but can be derived from the same underlying hydrolysis reaction.
Despite of the encouraging results, we would also like to indicate some difficulties related to the
definition and derivation of the parameters $\beta_{sub}$ and $\frac{k_{sub}}{K_m}$: First, we
extracted the formula for $\frac{k_{sub}}{K_m}$ using the expression for $\beta_{sub}$ given in
Eq.~\ref{betasub2}. This approach is problematic because the definition of $\beta_{sub}$ involves
the assumption of a well stirred cGMP concentration during a photoresponse, which is not very
accurate (see Fig.~\ref{fig_cGMP} and
\cite{DiBenedetto2005,DiBenedetto2003}). Second, if diffusion limits the rate of cGMP hydrolysis in
the physiological range, the experimentally observed value for $\frac{k_{sub}}{K_m}$ does not
reflect an intrinsic property of the chemical reaction. Instead, it depends strongly on diffusional
and geometrical details, and, therefore, on the experimental setup. For example, measurements of
the Michaelis constant $K_m$ were performed using fragments of disrupted rod outer segments with a
length only a fraction of the intact outer segment length
\cite{Leskovetal2000,Dumkeetal1995}. Eq.~\ref{eqforC} shows that the rate of cGMP hydrolysis
increases with decreasing fragment length $L$ (since $N_c\sim L$). Thus, the apparent value of the
Michaelis constant $K_m$ ($k_{sub}$ is assumed to be a true constant) that is needed to fit the
rate of cGMP hydrolysis will be higher in a suspension containing large fragments compared to a
suspension with small fragments, as it has been observed \cite{Leskovetal2000,Dumkeetal1995}.

In this work we have assumed that cGMP hydrolysis in the physiological range is diffusion limited,
and is independent of whether PDE is spontaneously- or light-activated. The agreement between our
theoretical results and experimental measurements indicates that the large disparity between
$\beta_{sub}$ and $\beta_{d}$ is largely due to their definition, and not due to biochemical
differences. For example, in \cite{DiBenedetto2003} the effect of spontaneously- and
light-activated PDE was modeled using two very different rates $k=0.042\mu M^{-1}s^{-1}$ and
$k^*=110\mu M^{-1}s^{-1}$. We will show now that the large discrepancy between $k$ and $k^*$ in
\cite{DiBenedetto2003} essentially originates from modeling needs. Indeed, cGMP hydrolysis by
spontaneously activated PDE was modeled as $k [PDE]_\sigma [cGMP]$, where $[PDE]_\sigma$ is the
surface concentration of PDE. In contrast, hydrolysis by light-activated PDE was modeled as $k^*
[PDE^*]_\sigma [cGMP]$, with $[PDE^*]_\sigma$ as the surface concentration of light-activated PDE.
By introducing the mean surface concentration of spontaneously activated PDE, $[PDE_s^*]_\sigma =
[PDE]_\sigma \frac{\mu_+}{\mu_-}$, we rewrite $k [PDE]_\sigma [cGMP]$ as $k \frac{\mu_-}{\mu_+}
[PDE_s^*]_\sigma [cGMP]$, which now has the same form as $k^* [PDE^*]_\sigma [cGMP]$. Inserting the
values $\mu_{+}=4\times 10^{-4}s^{-1}$ and $\mu_{-}=1.8 s^{-1}$ found in \cite{RiekeBaylor1996}, we
obtain $k \frac{\mu_-}{\mu_+} = 189\mu M^{-1}s^{-1}$, which is now comparable to $k^*=110\mu
M^{-1}s^{-1}$. We conclude that modeling cGMP hydrolysis by spontaneously- and light-activated PDE
in a similar way involves comparable parameters, indicating that hydrolysis may be indeed
independent of whether PDE is spontaneously- or light-activated.

%
%%%%%%%%%%%%%%%%%%%%%%%%%%%%%%%%%%%%%%%%%%%%%%%%%
\section{cGMP hydrolysis in cones}
%%%%%%%%%%%%%%%%%%%%%%%%%%%%%%%%%%%%%%%%%%%%%%%%
%
After having discussed in detail cGMP hydrolysis in rods, we now briefly explore hydrolysis in
cones. Similar to \cite{HolcmanKorenbrot2004}, our analysis for rods can be adapted to cones.
Unlike rods, cones do not contain disc in the outer segment. However, the membrane invaginations in
cones can be modeled similarly to discs in rods. Since the radius of the cone outer segment
decreases from the bottom versus the top, we can adapt our formulas to cones by replacing the disc
radius $R$ with a compartment dependent radius $R_n$. Thus, in cones, the rates $\tilde k_{1}$ and
$\tilde k_2$ depend on the compartment $n$. Therefore, the response to a photon absorption in cones
varies on the location where the photon has been absorbed. Since $k_{1} P^*_{s,c} $ depends
logarithmically on the compartment radius $R_n$ (see Eq.~\ref{Eqforbetad}), we suggest that the
value for the dark hydrolysis rate $\beta_d$ in cones should be of the same magnitude as found in
rods, see also \cite{HolcmanKorenbrot2005}.

%
%%%%%%%%%%%%%%%%%%%%%%%%%%%%%%%%%%%%%%%%%%
\section{Summary and Discussion}
%%%%%%%%%%%%%%%%%%%%%%%%%%%%%%%%%%%%%%%%%%%%
%

In this paper, we have studied the rate constant of cGMP hydrolysis by activated PDE in rod and
cone photoreceptors. Our analysis is based on the assumption that cGMP hydrolysis is diffusion
limited and determined by the encounter rate between cGMP and activated PDE. We derived an explicit
formula for the rate constant of cGMP hydrolysis by a single activated PDE molecule as a function
of the confined outer segment geometry and the cGMP diffusion constant (Eq.~\ref{fluxJa2}). Our
calculation takes into account the complex structure of the rod outer segment, uniformly divided by
a stack of parallel discs into homogenous microdomains, called compartments, and coupled to each
other via cGMP diffusion. We obtained analytical expressions for the rate constants $\beta_d$ and
$\beta_{sub}$. In addition, we give a set of effective equations that allow to model the
transversally well stirred cGMP concentration after a photon absorption.

Interestingly, we found that only the amount of spontaneously activated PDE in a single compartment
is needed to calculate the dark hydrolysis rate $\beta_d$ (see Eq.~\ref{Eqforbetad}). This result
differs from \cite{RiekeBaylor1996}, where the compartmentalization was not considered, and all
spontaneously activated PDE in the outer segment additively contribute to $\beta_d$. Because the
number of spontaneously activated PDE in the outer segment is by a factor $N_c\sim 10^3$ larger
compared to a single compartment, the catalytic activity of an excited PDE in
\cite{RiekeBaylor1996} was estimated much lower compared to what we found here. We computed the
PDE activity (given by the rate $\tilde k_1$ in Eq.~\ref{rates_exact}) to be around $1s^{-1}$,
whereas in \cite{RiekeBaylor1996} it is around $10^{-5}s^{-1}$. Using the rates for spontaneous PDE
activation and deactivation \cite{RiekeBaylor1996}, we estimate that the average number of
spontaneously activated PDE molecules in a single compartment is around one. Together with our
result for the PDE activity, this naturally explains the experimental value $\beta_d
\sim 1s^{-1}$. For the derivation of $\beta_d$, it was essential to assume that cGMP hydrolysis
occurs locally at the activated PDE site. In contrast, if cGMP hydrolysis occurred uniformly over
the disc surface, then the experimental value for $\beta_d$ could not be recovered without
introducing additional adjusting parameters (see appendix \ref{uniformcGMPHydrolysis}).

We have derived a set of equations (Eqs.~\ref{GcwithPh} and \ref{J_R2}) that allow to calculate the
time course of the transversally well stirred cGMP concentration following a photon absorption.
These equations model cGMP hydrolysis by spontaneously and light-activated PDE in a similar way.{
Under the assumption that cGMP concentration in the outer segment is well stirred, we derived an
expression for the rate $\beta_{sub}$ (Eq.~\ref{betasub},\ref{betasub2}), and the ratio
$\frac{k_{sub}}{K_m }$ (Eq.~\ref{ksubKm}).

Eq.~\ref{betasub} connects $\beta_{sub}$ to $\beta_d$ and gives a direct explanation why
$\beta_{sub}$ is found to be so much smaller than $\beta_d$. Our result suggests that the large
discrepancy between $\beta_d$ and $\beta_{sub}$ is largely due to their definitions: $\beta_d$
incorporates the effect of all spontaneously activated PDE in the outer segment, while
$\beta_{sub}$ accounts for only a single light-activated PDE.}

\newpage

\section*{APPENDIX}

\appendix
\section{Mean time to hydrolysis in a compartment}
\label{meanHydrolysisTime}

\begin{figure}[ht!]
\vspace{0cm}
  \begin{center}
    \mbox{ \includegraphics[scale=0.9]{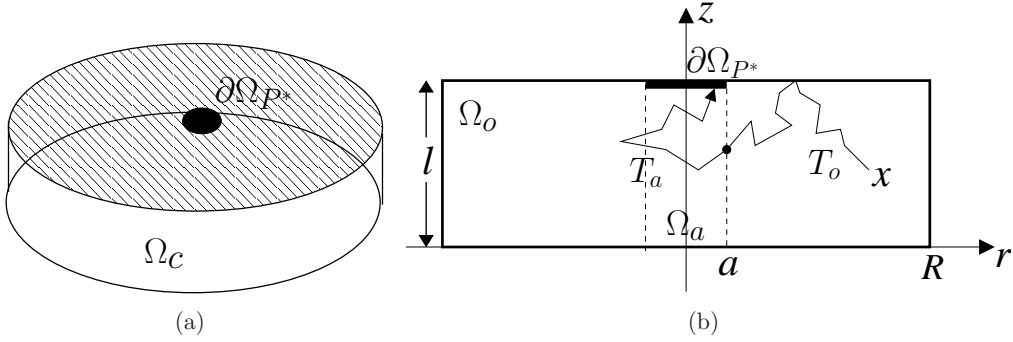} }
    \vspace{0cm}
    \caption{%\doublespacing
    (a) Cylindrical compartment $\Omega_c$ with an activated PDE molecule located centrally on the upper surface.
    (b) The  first time a cGMP molecule hits $\p \Omega_{P^*}$, when starting at position $\x$ in $\Omega_o$, is
    given by the sum of two times. First, the time $T_o$ for a molecule starting at $\x$ to arrive at the boundary
    $\partial \Omega_a$, and second, the time $T_a$ for the particle starting
    at $\partial \Omega_a$ to arrive at $\p \Omega_{P^*}$.}
    \label{FPT}
  \end{center}
\end{figure}

In this part of the appendix, we shall obtain a precise estimate for the mean time $\tau$ a cGMP
molecule starting uniformly distributed inside the cylindrical compartment $\Omega_c$ reaches the
activated PDE molecule, defined as the small surface patch $\p \Omega_{P^*}$ with radius $r=a$ (see
Fig~\ref{FPT}). The motion of the cGMP molecule is Brownian in $\Omega_c$. It is reflected all over
the boundary except at $\partial \Omega_{P^*}$, where is is absorbed. We denote by $T^{(x)}$ the
random initial time a cGMP molecule starting at $\x \in \Omega_c$ hits $\partial
\Omega_{P^*}$. Due to rotational invariance, the Mean First Passage Time (MFPT) $\tau(\x)=\mbox{E}[
T^{(x)}|\x(0)=\x]$ depends only on $r$ and $z$. Using a cylindrical coordinate system
$\x=(r,\varphi,z)$), we decompose the domain $\Omega_c$ into the inner cylinder
\beq
\Omega_{a}=\{\x \in \Omega_c| r\le a \}\,,
\eeq
and the hollow cylinder
\beq
\Omega_{o}=\Omega_c - \Omega_{a}= \{\x | a \le r \le R\}.
\eeq
We define the mean time $\tau$ as the average over a uniform initial distribution in $\Omega_c$,
\bea
\tau = \frac{1}{|\Omega_c|} \int_{\Omega_c} \tau(r,z) dV\,. \label{deftau}
\eea
Using that $\Omega_c = \Omega_o+\Omega_a$, we can rewrite Eq.~\ref{deftau} as
\bea
\tau &=&  \frac{|\Omega_c|-|\Omega_a|}{|\Omega_c|} \frac{1}{|\Omega_o|} \int_{\Omega_o} \tau(r,z) dV
+ \frac{|\Omega_c|-|\Omega_o|}{|\Omega_c|} \frac{1}{|\Omega_a|} \int_{\Omega_a} \tau(r,z) dV \nn\\
&=& \frac{1}{|\Omega_o|}\int_{\Omega_o} \tau(r,z) dV  -
 \frac{|\Omega_a|}{|\Omega_c|} \(
\frac{1}{|\Omega_o|}\int_{\Omega_o} \tau(r,z) dV -
\frac{1}{|\Omega_a|} \int_{\Omega_a} \tau(r,z) dV  \)\,, \label{deftau2_1}
\eea
where $\frac{1}{|\Omega_o|}\int_{\Omega_o} \tau(r,z) dV$ is the mean time for particles starting
uniformly distributed in $\Omega_o$ to hit $\partial \Omega_{P^*}$, and
$\frac{1}{|\Omega_a|}\int_{\Omega_a} \tau(r,z) dV$ is the mean time to hit $\partial
\Omega_{P^*}$ for particles starting uniformly distributed in $\Omega_a$. Because particles
originating from $\Omega_{o}$ have to reach $\Omega_{a}$ before hitting $\partial
\Omega_{P^*}$, it is plausible (and can also be shown) that the mean time to $\partial
\Omega_{P^*}$ for particles starting uniformly in $\Omega_{o}$ is larger than the mean time for
particles starting uniformly in $\Omega_{a}$. Furthermore, for $a \ll R$, we have $|\Omega_a|
\ll |\Omega_c|$. From this, we finally obtain
\bea
\tau =  \frac{1}{|\Omega_c|} \int_{\Omega_c} \tau(r,z) dV
\approx  \frac{1}{|\Omega_o|}\int_{\Omega_o} \tau(r,z) dV
\,.\label{deftau2}
\eea
Thus, for $a \ll R$, the mean time $\tau$ is well approximated  by the mean time for particles
starting uniformly in $\Omega_o$ to hit $\partial \Omega_{P^*}$.

We shall now estimate $\tau(r,z)$, by considering the equation
\cite{BookSchuss,BookGardiner}
\bea
D_{cG}\( \frac{\p^2}{\p r^2}+ \frac{1}{r} \frac{\p}{\p r}
+\frac{\p^2}{\p z^2}\)\tau (r,z)\,, &=&-1
\quad  0<z<l\,, 0\le r < R \label{unscaledEqtau}\\
\tau(r,z)&=& 0\,, \quad  z=l \,, r < a \nn \\
\frac{\partial}{\partial z}\tau (r,z)&=&0\,,  \quad z=l \,,  r> a  \nn \\
\frac{\partial}{\partial z}\tau (r,z)&=&0\,,  \quad z=0  \nn \\
\frac{\partial}{\partial r}\tau (r,z) &=&0\,, \quad  r=R \,. \nn
\eea\
We will first estimate the average time
\bea
\tau(r)= \frac{1}{l} \int_0^{l} \tau(r,z) dz \label{deftau(r)}\,.
\eea
Using the dimensionless variables
\bea
x=\frac{r}{a}\,, \quad y=\frac{z}{a}\,, \quad
\hat \tau(x,y) =\frac{D_{cG}}{R^2} \tau(r,z)\,, \quad
\alpha=\frac{a}{R}\,, \quad  \beta=\frac{l}{a}, \quad  x_\alpha=\frac{1}{\alpha}
\label{defscaling}\,,
\eea
equation \ref{unscaledEqtau} becomes
\bea
\( \frac{1}{x} \frac{\p}{\p x} x\frac{\p}{\p x} +
\frac{\p^2}{\p y^2}\) \hat \tau (x,y) &=&-\alpha^2\,,
\quad  0<y<\beta\,,\, 0\le x <  x_\alpha \label{scaledEqtau}\\
\hat \tau(x,y)&=& 0\,, \quad  y=\beta\,, \, x < 1 \nn \\
\frac{\partial}{\partial y}\hat \tau (x,y)&=&0\,,  \quad  y=\beta \,,  x> 1  \nn \\
\frac{\partial}{\partial y}\hat \tau (x,y)&=&0\,,  \quad  y=0  \nn \\
\frac{\partial}{\partial x}\hat \tau (x,y) &=&0\,, \quad  x=x_\alpha \,, \nn
\eea\
and  Eq.~\ref{deftau(r)}
\bea
\hat \tau(x)= \frac{1}{\beta} \int_0^{\beta} \hat \tau(x,y) dy\,. \label{defT(x)}
\eea
We integrate Eq.~\ref{scaledEqtau} over the variable $y$ to derive an equation for $\hat\tau(x)$
for $x\ge 1$. Taking into account the boundary conditions at $y=0$ and $y=\beta$, we obtain
\bea
\frac{1}{x} \frac{\p}{\p x} x\frac{\p}{\p x} \hat \tau(x) &=& -\alpha^2 \,, \quad x>1 \label{diffEqF}\\
\frac{\p}{\p x} \hat \tau(x) &=& 0 \quad \mbox{for } x=x_\alpha \nn
\eea
The solution is given
\bea\label{T1}
\hat \tau(x)= f(\alpha,\beta) + \frac{1}{2}\ln (x) -\frac{1}{4}\alpha^2 (x^2 -1)\,,
\eea
with
\bea
 f(\alpha,\beta)= \hat \tau(1) =\frac{1}{\beta} \int_0^{\beta} \hat \tau(1,y) dy\,.
\eea
Hence, for $r\ge a$, we have
\bea\label{tau(r)}
\tau(r)&=& \frac{R^2}{D_{cG}}f(\alpha,\beta) +
 \frac{R^2}{D_{cG}} \(\frac{1}{2}\ln (x) -\frac{1}{4}\alpha^2 (x^2 -1) \) \\
&=& \tau_a + \tau_o(r) \,,
\eea
where we defined
\bea
\tau_a &=&  \frac{R^2}{D_{cG}}f(\alpha,\beta) \label{meanTimeTaua}\,, \\
\tau_{o}(r) &=& \frac{R^2}{D_{cG}}\(\frac{1}{2}\ln \(\frac{r}{a}\)
- \frac{1}{4}\frac{r^2-a^2}{R^2} \)\,. \label{meanTimeTauo(r)}
\eea
Eq.~\ref{tau(r)} has an intuitive interpretation: the mean time $\tau(r)$ for a cGMP molecule,
uniformly distributed at  $r>a$, is the sum of the mean time $\tau_o(r)$ to the boundary $r=a$ plus
the mean time $\tau_a$ from the surface $\p \Omega_a$ to go to $\partial
\Omega_{P^*}$  (see Fig~\ref{FPT}).

\begin{figure}[ht!]
\vspace{0cm}
  \begin{center}
    \mbox{ \includegraphics[scale=1]{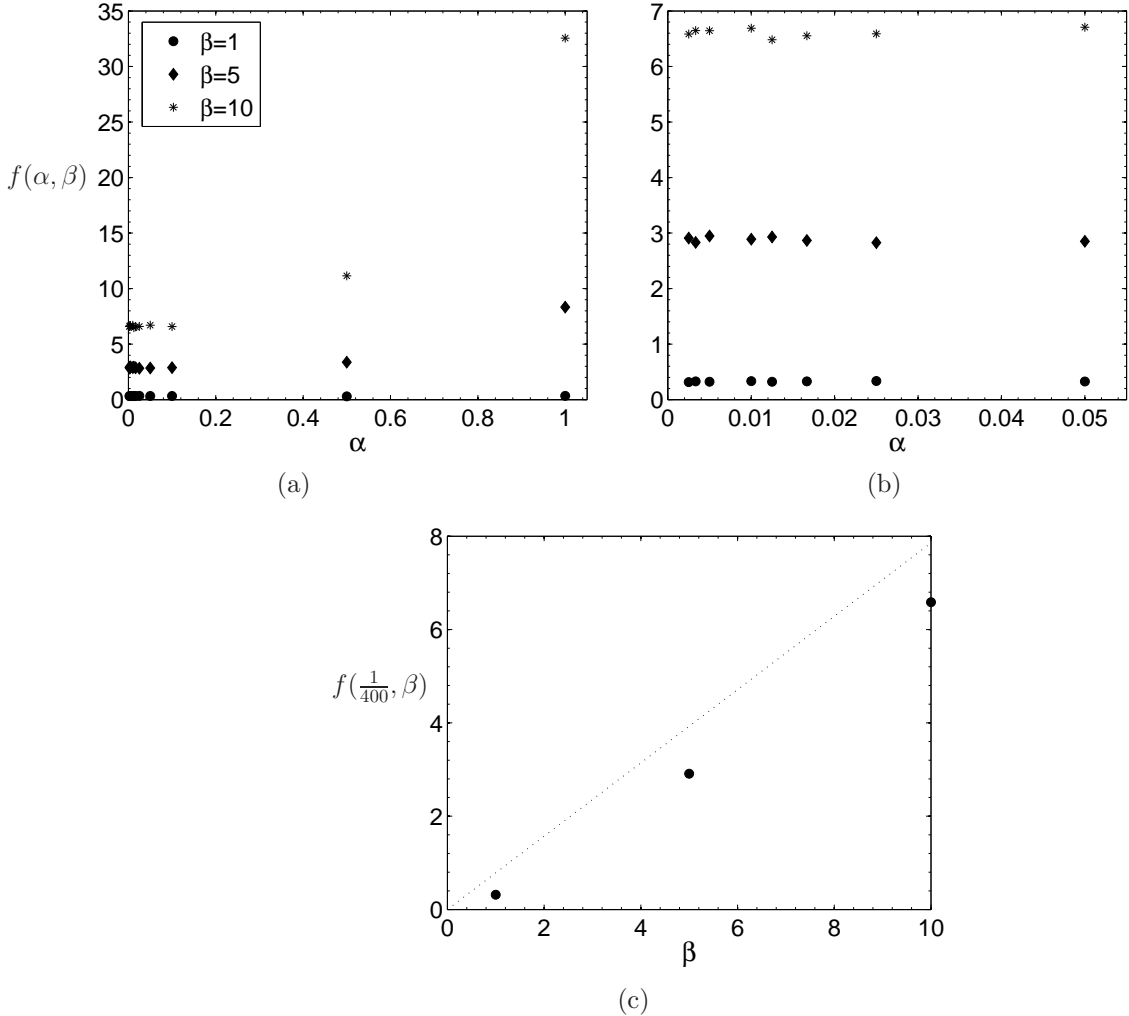} }
    \vspace{0cm}
    \caption{%\doublespacing
    Numerical evaluation of the function $f(\alpha,\beta)$ ($\alpha=a/R$ and $\beta=l/a$).
    Each data point is obtained using the Brownian simulation of 10000 cGMP molecules.
    (a) $f(\alpha,\beta)$ for different values $\alpha$ and $\beta$.
    For $\alpha=1$ we have $f(1,\beta)=\frac{\beta^2}{3}$ (Eq.~\ref{asymp_f1}).
    (b) Same data as in (a), restricted to small $\alpha$.
    (c) Plot of $f(\frac{1}{400},\beta)\approx g(\beta)$ (same data as in (a) and (b)).
    The dashed curve represents the asymptotic $\frac{\pi}{4}\beta$ (Eq.~\ref{beta}),
    achieved for $\alpha \to 0$ and $\beta\to \infty$ .
    For values $\beta \le 10$ used in the simulations,
    the behavior of $g(\beta)$ is close, but not yet in full agreement with $\frac{\pi}{4}\beta$.
     }
     \label{fig_Taua}
  \end{center}
\end{figure}

By averaging over a uniform initial distribution $\rho =\frac{1}{\pi(R^2-a^2)}$ in $\Omega_o$, the
overall mean time $\tau$ in Eq.~\ref{deftau2} is given by
\bea
\tau &=& \frac{1}{|\Omega_o|}\int_{\Omega_o} \tau(r,z) dV = 2\pi \rho \int \tau(r) r dr\nn \\
&=&  \frac{R^2}{D_{cG}}f(\alpha,\beta) + \frac{R^2}{8 D_{cG}}\frac{-4 \ln\(\alpha \) - 3 + 4
\alpha^2 - \alpha^4 }{1-\alpha^2} \\
&=& \tau_a +\tau_o \,, \label{resTauoExact}
\eea
where we defined $\tau_o$ as
\bea
\tau_o = 2\pi \rho \int \tau_o (r) r dr = \frac{R^2}{8 D_{cG}}\frac{-4 \ln\(\alpha \) - 3 + 4
\alpha^2 - \alpha^4 }{1-\alpha^2}.
\eea
The leading order expansion of $\tau_{o}$ for $\alpha  \ll 1$ is
\bea\label{meantime}
\tau_{o} = \frac{R^2}{D_{cG}}\(\frac{1}{2} \ln\(\frac{R}{a}\) - \frac{3}{8}\) \,.
\eea
%We remark that the leading order estimation in Eq.~\ref{meantime} is
%also valid for an initial distribution that corresponds to the
%concentration profile in Eq.~\ref{equConc}, and also for a $P^*$
%molecule that is located outside the center \cite{HolcmanetalNE3}.
%
%Note that the limit $\alpha  \to  0$ with constant $\beta=l/a$ can only be obtained from $R
%\to \infty$ and not from $a\to 0$. It follows from Eq.~\ref{scaledEqtau} that $f(\alpha,\beta)$
%remains finite as $\alpha \to 0$. Thus,
For $\alpha \ll 1$ (see also Fig.~\ref{fig_Taua}b), we can approximate $f(\alpha,\beta)$ by
$f(0,\beta)=g(\beta)$ and obtain ($\beta=\frac{l}{a}$)
\bea\label{asumtoticf}
\tau_a & \approx& \frac{R^2}{D_{cG}}g\(\frac{l}{a}\) \,, \quad \alpha \ll 1\,.
\eea
Altogether, for $\alpha  \ll 1$, the mean time $\tau$ in Eq.~\ref{deftau2} is given by
\bea \label{tauasymp}
\tau &=&  \tau_a +\tau_o \approx \frac{R^2}{D_{cG}}\[ g\(\frac{l}{a}\) + \frac{1}{2} \ln\(\frac{R}{a}\) -
\frac{3}{8}\]\,.
\eea
To derive an explicit expression for $f(\alpha,\beta)$ and $g(\beta)$ is a difficult mathematical
problem. Nonetheless, we shall obtain some asymptotic limits for $f(\alpha,\beta)$.  For $\beta \to
0$ (corresponding to $l\to 0$), we have $\tau_a \to 0$, and therefore $f(\alpha,0) = 0$. For $\beta
\to \infty$, the time $\tau_a$ diverges to infinity, and $f(\alpha,\beta)\to \infty $. For $a=R$,
corresponding to $\alpha=1$, we have $\tau_a=\frac{l^2}{3D_{cG}}$, from which it follows that
\bea
f(1,\beta)=\frac{\beta^2}{3}\,. \label{asymp_f1}
\eea
Finally, the small hole theory \cite{HolcmanetalNE1} predicts that when $l \sim R \ll a$ (which
implies that $\alpha\sim \beta$), the mean time $\tau$ is asymptotically given by
\bea\label{small hole}
\tau \approx \frac{V}{4 D_{cG} a}= \frac{\pi R^2}{4 D_{cG}}\beta \,.
\eea
By comparing Eq.~\ref{small hole} with Eq.~\ref{tauasymp} ($\ln(\alpha)$ can be neglected compared
to $\beta$ for $\alpha\sim \beta$), we obtain the asymptotic
\bea \label{beta}
g(\beta)  \sim \frac{\pi}{4}\beta\,, \quad  \beta\gg 1.
\eea
So far, we have only an asymptotic expansion for $\beta\gg 1$. To explore a much larger parameter
space, we decided to run Brownian simulations (10000 cGMP molecules are initially uniformly
distributed over the lateral surface of  $\partial \Omega_{a}$) to estimate $\tau_{a}$ and
$f(\alpha,\beta)$. The numerical results for $f(\alpha,\beta)$ are summarized in
Fig.~\ref{fig_Taua}. Fig.~\ref{fig_Taua}b shows that $f(\alpha,\beta)$ can be well approximated by
$f(0,\beta)=g(\beta)$ for $\alpha \lesssim 0.05$.

\cleardoublepage
\section{Model with uniform hydrolysis on the disc surfaces}
\label{uniformcGMPHydrolysis}
In this section, we consider a model that is based on cGMP hydrolysis occurring uniformly on the
disc surface (see also \cite{DiBenedetto2003,DiBenedetto2005}). This is very different from the
situation presented within the main body of the paper, where we assumed that cGMP hydrolysis occurs
locally at the $P^*$ site. We show now that uniform cGMP hydrolysis leads to a dark rate constant
proportional to ${D_{cG}}/{l^2}\sim 10^5s^{-1}$, which is very different from Eq.~\ref{Eqforbetad}
(see also section \ref{compExpRes}). Thus, in order to account for the experimental value
$\beta_d\sim 1s^{-1}$, one is forced to introduce a small adapting parameter $\kappa_h$.

We will analyze two different scenarios: In one situation, synthesis and hydrolysis of cGMP are
both modelled by boundary source terms. In another, only hydrolysis is modelled by a boundary
source term, while synthesis occurs uniformly within the cytoplasmic volume.
\subsection{Model with boundary source terms for hydrolysis and synthesis}
The reaction-diffusion equation for cGMP concentration $C(z,r,t)$ inside a compartment is given by
\bea
\frac{\partial }{\partial t}C(z,r,t) &=&  D_{cG} \Delta C(z,r,t)\,,\label{appDiffEq}\\
-D_{cG} \frac{\partial C(z,r,t)} {\partial z}\Big |_{\overset{z=0}{z=l}} &=& -\kappa_h
C(z,r,t)\Big |_{\overset{z=0}{z=l}}
+ \alpha_\sigma(t)  \label{appBoundCond1} \\
 D_{cG} \frac{\partial C(z,r,t)} {\partial r}\big |_{r=R} &=& 0
\eea
The steady state expressions for the concentration $C$ of Eq.~\ref{appDiffEq} and the hydrolysis
rate $J_h$ are ($G_c=\pi R^2 C$)
\bea
C&=&\frac{\alpha_\sigma}{\kappa_h} \,,\\
J_h&=&2\pi R^2 \kappa_h C = 2 \frac{\kappa_h }{l} G_c\label{app_Jh1}\,.
\eea
Because hydrolysis and synthesis are both modelled by fluxes originating from the same boundary,
the equilibrium only reflects the balance of the fluxes, and does not involve cGMP diffusion.

\subsection{Model with boundary source term for hydrolysis, and a volume synthesis rate}
To include cGMP diffusion (see section
\ref{AnalysiscGMPhydrolysis}), we now model cGMP synthesis by a uniform volume production rate
$\alpha_v(t)= 2\alpha_\sigma(t)/l$. This model avoids the problems arising when cGMP synthesis and
hydrolysis are both modelled by surface fluxes originating from the same boundary. The equation for
the cGMP concentration reads
\bea\label{appDiffEq3}
\frac{\partial}{\partial t}C(z,r,t) &=&  D_{cG} \Delta C(z,r,t) + \alpha_v(t)\,, \label{appDiffEq2}\\
-D_{cG} \frac{\partial C(z,r,t)} {\partial z}\Big |_{\overset{z=0}{z=l}} &=& -\kappa_h  C(z,r,t) \Big |_{\overset{z=0}{z=l}}\\
 D_{cG} \frac{\partial C(z,r,t)} {\partial r}\Big |_{r=R} &=& 0
\eea
The steady state solution of Eq.~\ref{appDiffEq3} is
\bea
C(z)=\frac{\hat C}{2} \frac{z}{l} (1-\frac{z}{l}) + \frac{ \hat C }{2 \beta}\,,
\eea
where we introduced the parameter $\beta$ and the concentration $\hat C$ as
\bea\label{appdimless}
\beta = \frac{\kappa_h l}{D_{cG}} \,, \quad
\hat C = \alpha_v  \frac{l^2}{D_{cG}}\,.
\eea
At equilibrium, the number of cGMP molecules $G_c$ and the
hydrolysis rate $J_h$ in a compartment are
\bea
G_c &=&  \pi R^2 \int_0^l C(z) dz = \hat C \(\frac{1}{12}+\frac{ 1}{2 \beta}\)  \pi R^2 l \,,\\
J_{h} &=&  \pi R^2 D_{cG}\(  \frac{d C(z)}{dz }\big |_{z=0} -
\frac{d C(z)} {dz }\big |_{z=1}\) = \pi R^2 l \alpha_v
 \nn\\
%&=& 2  \pi R^2\kappa_h P^*_\sigma C(0) =  \frac{ D_{cG}}{l^2} \pi R^2 l \hat  C  \nn\\
&=& \frac{ D_{cG}}{l^2} \(\frac{1}{12}+
\frac{1}{2 \beta}\)^{-1} G_c \,. \label{appflux}
\eea
Contrary to Eq.~\ref{app_Jh1}, the flux $J_h$ in Eq.~\ref{appflux} depends on cGMP diffusion
constant. In the limit $\beta \to \infty$ ($\kappa_h \to \infty$) (perfectly absorbing boundaries),
we obtain $J_h=G_c/\tau$, where $\tau=\frac{l^2}{12D_{cG}}$ is the mean time for a molecule to
reach the boundaries at $z=l$ or $z=0$. On the other hand, in the limit $\beta\to 0$, we obtain
$J_h=\frac{D_{cG}}{l^2}2\beta G_c=2 \frac{\kappa_h}{l} G_c$. Thus, we recover the expression given
in Eq.~\ref{app_Jh1}.

Eq.~\ref{appflux} is formally equivalent to Eq.~\ref{fluxJa2}, however, the physical content is
very different. $J_h$ in Eq.~\ref{appflux} is proportional to the rate by which cGMP molecules
collide with the disc surfaces, which is of the order $D_{cG}/l^2\sim 10^5 s^{-1}$. In contrast,
$J_h$ in Eq.~\ref{fluxJa2} is determined by the rate by which cGMP molecules find $P^*$, given by
$D_{cG}/R^2\sim 1s^{-1}$. In order to obtain $\beta_d \sim 1s^{-1}$ from Eq.~\ref{appflux}, one
needs a small value for the adapting parameter $\kappa_h$.

\subsection*{Acknowledgments}
The authors would like to thank Maria Corado for carefully reading the manuscript. J.R. thanks the
FRM-foundation for support.

%-----------------------------------REFERENCES ----------------------------------------------

%\bibliographystyle{apsrev}
%\bibliography{Bib_Phototransduction,Bib_ReactionTheoryDiffusionNarrowEscape,Bib_Books}

\begin{thebibliography}{50}
\expandafter\ifx\csname natexlab\endcsname\relax\def\natexlab#1{#1}\fi
\expandafter\ifx\csname bibnamefont\endcsname\relax
  \def\bibnamefont#1{#1}\fi
\expandafter\ifx\csname bibfnamefont\endcsname\relax
  \def\bibfnamefont#1{#1}\fi
\expandafter\ifx\csname citenamefont\endcsname\relax
  \def\citenamefont#1{#1}\fi
\expandafter\ifx\csname url\endcsname\relax
  \def\url#1{\texttt{#1}}\fi
\expandafter\ifx\csname urlprefix\endcsname\relax\def\urlprefix{URL }\fi
\providecommand{\bibinfo}[2]{#2}
\providecommand{\eprint}[2][]{\url{#2}}

\bibitem[{\citenamefont{Arrhenius}(1889)}]{Arrhenius1889}
\bibinfo{author}{\bibfnamefont{S.}~\bibnamefont{Arrhenius}},
  \bibinfo{journal}{Z. Phys. Chem.} \textbf{\bibinfo{volume}{4}},
  \bibinfo{pages}{226} (\bibinfo{year}{1889}).

\bibitem[{\citenamefont{Kramers}(1940)}]{Kramers1940}
\bibinfo{author}{\bibfnamefont{H.~A.} \bibnamefont{Kramers}},
  \bibinfo{journal}{Physica (Amsterdam)} \textbf{\bibinfo{volume}{7}},
  \bibinfo{pages}{284} (\bibinfo{year}{1940}).

\bibitem[{\citenamefont{H\"anggi et~al.}(1990)\citenamefont{H\"anggi, Talkner,
  and Borkovec}}]{Haenggietal1990}
\bibinfo{author}{\bibfnamefont{P.}~\bibnamefont{H\"anggi}},
  \bibinfo{author}{\bibfnamefont{P.}~\bibnamefont{Talkner}}, \bibnamefont{and}
  \bibinfo{author}{\bibfnamefont{M.}~\bibnamefont{Borkovec}},
  \bibinfo{journal}{Rev. Mod. Physics} \textbf{\bibinfo{volume}{62}},
  \bibinfo{pages}{251} (\bibinfo{year}{1990}).

\bibitem[{\citenamefont{Schuss}(1980)}]{BookSchuss}
\bibinfo{author}{\bibfnamefont{Z.}~\bibnamefont{Schuss}},
  \emph{\bibinfo{title}{Theory and Applications of Stochastic Differential
  Equations}} (\bibinfo{publisher}{Wiley Series in Probability and Statistics,
  John Wiley Sons, Inc., New York}, \bibinfo{year}{1980}).

\bibitem[{\citenamefont{von Smoluchowski}(1914)}]{Smoluchowski1914}
\bibinfo{author}{\bibfnamefont{M.}~\bibnamefont{von Smoluchowski}},
  \bibinfo{journal}{Wien Berlin} \textbf{\bibinfo{volume}{123}},
  \bibinfo{pages}{12381} (\bibinfo{year}{1914}).

\bibitem[{\citenamefont{Berg and Purcell}(1977)}]{BergPurcell1977}
\bibinfo{author}{\bibfnamefont{H.~C.} \bibnamefont{Berg}} \bibnamefont{and}
  \bibinfo{author}{\bibfnamefont{M.}~\bibnamefont{Purcell}},
  \bibinfo{journal}{Biophys. J.} \textbf{\bibinfo{volume}{20}},
  \bibinfo{pages}{193} (\bibinfo{year}{1977}).

\bibitem[{\citenamefont{Zwanzig}(1990)}]{ZwanzigPNAS1990}
\bibinfo{author}{\bibfnamefont{R.}~\bibnamefont{Zwanzig}},
  \bibinfo{journal}{Proc. Natl. Acad. Sci. USA} \textbf{\bibinfo{volume}{87}},
  \bibinfo{pages}{5856} (\bibinfo{year}{1990}).

\bibitem[{\citenamefont{Szabo et~al.}(1980)\citenamefont{Szabo, Schulten, and
  Schulten}}]{SzaboSchultenSchulten1980}
\bibinfo{author}{\bibfnamefont{A.}~\bibnamefont{Szabo}},
  \bibinfo{author}{\bibfnamefont{K.}~\bibnamefont{Schulten}}, \bibnamefont{and}
  \bibinfo{author}{\bibfnamefont{Z.}~\bibnamefont{Schulten}},
  \bibinfo{journal}{J. Chem. Phys.} \textbf{\bibinfo{volume}{72}},
  \bibinfo{pages}{4350} (\bibinfo{year}{1980}).

\bibitem[{\citenamefont{Schulten et~al.}(1981)\citenamefont{Schulten, Schulten,
  and Szabo}}]{SzaboSchultenSchulten1981}
\bibinfo{author}{\bibfnamefont{K.}~\bibnamefont{Schulten}},
  \bibinfo{author}{\bibfnamefont{Z.}~\bibnamefont{Schulten}}, \bibnamefont{and}
  \bibinfo{author}{\bibfnamefont{A.}~\bibnamefont{Szabo}}, \bibinfo{journal}{J.
  Chem. Phys.} \textbf{\bibinfo{volume}{72}}, \bibinfo{pages}{4426}
  (\bibinfo{year}{1981}).

\bibitem[{\citenamefont{Perico and Battezzati}(1981)}]{PericoBattezzati1981}
\bibinfo{author}{\bibfnamefont{A.}~\bibnamefont{Perico}} \bibnamefont{and}
  \bibinfo{author}{\bibfnamefont{M.}~\bibnamefont{Battezzati}},
  \bibinfo{journal}{J. Chem. Phys.} \textbf{\bibinfo{volume}{75}},
  \bibinfo{pages}{4430} (\bibinfo{year}{1981}).

\bibitem[{\citenamefont{Wilemski and Fixman}(1973)}]{WilemskiFixman1973}
\bibinfo{author}{\bibfnamefont{G.}~\bibnamefont{Wilemski}} \bibnamefont{and}
  \bibinfo{author}{\bibfnamefont{M.}~\bibnamefont{Fixman}},
  \bibinfo{journal}{J. Chem. Phys.} \textbf{\bibinfo{volume}{58}},
  \bibinfo{pages}{4009} (\bibinfo{year}{1973}).

\bibitem[{\citenamefont{Collins and Kimball}(1949)}]{CollinsKimball1949}
\bibinfo{author}{\bibfnamefont{F.~C.} \bibnamefont{Collins}} \bibnamefont{and}
  \bibinfo{author}{\bibfnamefont{G.~E.} \bibnamefont{Kimball}},
  \bibinfo{journal}{J. Colloid Sci.} \textbf{\bibinfo{volume}{4}},
  \bibinfo{pages}{425} (\bibinfo{year}{1949}).

\bibitem[{\citenamefont{Berezhkovskii et~al.}(2004)\citenamefont{Berezhkovskii,
  Makhnovskii, Monine, Zitserman, and Shvartsman}}]{Berezhkovskiietal2004}
\bibinfo{author}{\bibfnamefont{A.~M.} \bibnamefont{Berezhkovskii}},
  \bibinfo{author}{\bibfnamefont{Y.~A.} \bibnamefont{Makhnovskii}},
  \bibinfo{author}{\bibfnamefont{M.~I.} \bibnamefont{Monine}},
  \bibinfo{author}{\bibfnamefont{V.~Y.} \bibnamefont{Zitserman}},
  \bibnamefont{and} \bibinfo{author}{\bibfnamefont{S.~Y.}
  \bibnamefont{Shvartsman}}, \bibinfo{journal}{J.~Chem.~Phys.}
  \textbf{\bibinfo{volume}{121}}, \bibinfo{pages}{11390}
  (\bibinfo{year}{2004}).

\bibitem[{\citenamefont{Taflia and Holcman}(2007)}]{TafliaHolcman2007}
\bibinfo{author}{\bibfnamefont{A.}~\bibnamefont{Taflia}} \bibnamefont{and}
  \bibinfo{author}{\bibfnamefont{D.}~\bibnamefont{Holcman}},
  \bibinfo{journal}{J. Chem. Phys.} \textbf{\bibinfo{volume}{126}},
  \bibinfo{pages}{23407} (\bibinfo{year}{2007}).

\bibitem[{\citenamefont{Holcman and Schuss}(2005)}]{HolcmanSchuss2005_JCP}
\bibinfo{author}{\bibfnamefont{D.}~\bibnamefont{Holcman}} \bibnamefont{and}
  \bibinfo{author}{\bibfnamefont{Z.}~\bibnamefont{Schuss}},
  \bibinfo{journal}{J. Chemical Physics} \textbf{\bibinfo{volume}{122}},
  \bibinfo{pages}{114710} (\bibinfo{year}{2005}).

\bibitem[{\citenamefont{Grigoriev et~al.}(2002)\citenamefont{Grigoriev,
  Makhnovskii, Berezhkovskii, and Zitserman}}]{Grigorievetal2002}
\bibinfo{author}{\bibfnamefont{I.~V.} \bibnamefont{Grigoriev}},
  \bibinfo{author}{\bibfnamefont{Y.~A.} \bibnamefont{Makhnovskii}},
  \bibinfo{author}{\bibfnamefont{A.~M.} \bibnamefont{Berezhkovskii}},
  \bibnamefont{and} \bibinfo{author}{\bibfnamefont{V.~Y.}
  \bibnamefont{Zitserman}}, \bibinfo{journal}{J.~Chem.~Phys.}
  \textbf{\bibinfo{volume}{116}}, \bibinfo{pages}{9574} (\bibinfo{year}{2002}).

\bibitem[{\citenamefont{Schuss et~al.}(2007)\citenamefont{Schuss, Singer, and
  Holcman}}]{HolcmanPNAS2007}
\bibinfo{author}{\bibfnamefont{Z.}~\bibnamefont{Schuss}},
  \bibinfo{author}{\bibfnamefont{A.}~\bibnamefont{Singer}}, \bibnamefont{and}
  \bibinfo{author}{\bibfnamefont{D.}~\bibnamefont{Holcman}},
  \bibinfo{journal}{Proc. Natl. Acad. Sci. USA} \textbf{\bibinfo{volume}{104}},
  \bibinfo{pages}{16098} (\bibinfo{year}{2007}).

\bibitem[{\citenamefont{Hecht et~al.}(1942)\citenamefont{Hecht, Shlaer, and
  Pirenne}}]{Hechtetal1942}
\bibinfo{author}{\bibfnamefont{S.}~\bibnamefont{Hecht}},
  \bibinfo{author}{\bibfnamefont{S.}~\bibnamefont{Shlaer}}, \bibnamefont{and}
  \bibinfo{author}{\bibfnamefont{M.}~\bibnamefont{Pirenne}},
  \bibinfo{journal}{J.~Gen.~Physiol.} \textbf{\bibinfo{volume}{25}},
  \bibinfo{pages}{819} (\bibinfo{year}{1942}).

\bibitem[{\citenamefont{Sakitt}(1972)}]{Sakitt1972}
\bibinfo{author}{\bibfnamefont{B.}~\bibnamefont{Sakitt}},
  \bibinfo{journal}{J.~Physiol.} \textbf{\bibinfo{volume}{223}},
  \bibinfo{pages}{131} (\bibinfo{year}{1972}).

\bibitem[{\citenamefont{Baylor et~al.}(1979)\citenamefont{Baylor, Lamb, and
  Yau}}]{BaylorLambYau1979}
\bibinfo{author}{\bibfnamefont{D.}~\bibnamefont{Baylor}},
  \bibinfo{author}{\bibfnamefont{T.}~\bibnamefont{Lamb}}, \bibnamefont{and}
  \bibinfo{author}{\bibfnamefont{K.-W.} \bibnamefont{Yau}},
  \bibinfo{journal}{J. Physiol.} \textbf{\bibinfo{volume}{288}},
  \bibinfo{pages}{613} (\bibinfo{year}{1979}).

\bibitem[{\citenamefont{Rieke and Baylor}(1998)}]{RevRiekeBaylor1998}
\bibinfo{author}{\bibfnamefont{F.}~\bibnamefont{Rieke}} \bibnamefont{and}
  \bibinfo{author}{\bibfnamefont{D.}~\bibnamefont{Baylor}},
  \bibinfo{journal}{Rev. of Mod. Phys.} \textbf{\bibinfo{volume}{70}}
  (\bibinfo{year}{1998}).

\bibitem[{\citenamefont{Pugh~Jr and Lamb}(1993)}]{RevPughLamb1993}
\bibinfo{author}{\bibfnamefont{E.}~\bibnamefont{Pugh~Jr}} \bibnamefont{and}
  \bibinfo{author}{\bibfnamefont{T.}~\bibnamefont{Lamb}},
  \bibinfo{journal}{Biochim. et Biophys. Acta} \textbf{\bibinfo{volume}{1141}},
  \bibinfo{pages}{111} (\bibinfo{year}{1993}).

\bibitem[{\citenamefont{Pugh~Jr and Lamb}(2000)}]{RevPughLamb2000}
\bibinfo{author}{\bibfnamefont{E.}~\bibnamefont{Pugh~Jr}} \bibnamefont{and}
  \bibinfo{author}{\bibfnamefont{T.}~\bibnamefont{Lamb}},
  \bibinfo{journal}{Handbook of Biological Physics}
  \textbf{\bibinfo{volume}{3}} (\bibinfo{year}{2000}).

\bibitem[{\citenamefont{Burns and Baylor}(2001)}]{RevBurnsBaylor2001}
\bibinfo{author}{\bibfnamefont{M.}~\bibnamefont{Burns}} \bibnamefont{and}
  \bibinfo{author}{\bibfnamefont{D.}~\bibnamefont{Baylor}},
  \bibinfo{journal}{Annu. Rev. Neurosci.} \textbf{\bibinfo{volume}{24}},
  \bibinfo{pages}{779} (\bibinfo{year}{2001}).

\bibitem[{\citenamefont{Arshavsky et~al.}(2002)\citenamefont{Arshavsky, Lamb,
  and Pugh~Jr}}]{RevArshavskyLambPugh2002}
\bibinfo{author}{\bibfnamefont{V.}~\bibnamefont{Arshavsky}},
  \bibinfo{author}{\bibfnamefont{T.}~\bibnamefont{Lamb}}, \bibnamefont{and}
  \bibinfo{author}{\bibfnamefont{E.}~\bibnamefont{Pugh~Jr}},
  \bibinfo{journal}{Annu. Rev. Physiol.} \textbf{\bibinfo{volume}{64}},
  \bibinfo{pages}{153} (\bibinfo{year}{2002}).

\bibitem[{\citenamefont{Burns and Arshavsky}(2005)}]{RevBurnsArshavsky2005}
\bibinfo{author}{\bibfnamefont{M.}~\bibnamefont{Burns}} \bibnamefont{and}
  \bibinfo{author}{\bibfnamefont{V.}~\bibnamefont{Arshavsky}},
  \bibinfo{journal}{Neuron} \textbf{\bibinfo{volume}{48}}, \bibinfo{pages}{387}
  (\bibinfo{year}{2005}).

\bibitem[{\citenamefont{Andreucci et~al.}(2003)\citenamefont{Andreucci,
  Bisegna, Caruso, Hamm, and DiBenedetto}}]{DiBenedetto2003}
\bibinfo{author}{\bibfnamefont{D.}~\bibnamefont{Andreucci}},
  \bibinfo{author}{\bibfnamefont{P.}~\bibnamefont{Bisegna}},
  \bibinfo{author}{\bibfnamefont{G.}~\bibnamefont{Caruso}},
  \bibinfo{author}{\bibfnamefont{H.}~\bibnamefont{Hamm}}, \bibnamefont{and}
  \bibinfo{author}{\bibfnamefont{E.}~\bibnamefont{DiBenedetto}},
  \bibinfo{journal}{Biophys. J.} \textbf{\bibinfo{volume}{85}},
  \bibinfo{pages}{1358} (\bibinfo{year}{2003}).

\bibitem[{\citenamefont{Caruso et~al.}(2006)\citenamefont{Caruso, Bisegna,
  Shen, Andreucci, Hamm, and DiBenedetto}}]{DiBenedetto2006}
\bibinfo{author}{\bibfnamefont{G.}~\bibnamefont{Caruso}},
  \bibinfo{author}{\bibfnamefont{P.}~\bibnamefont{Bisegna}},
  \bibinfo{author}{\bibfnamefont{L.}~\bibnamefont{Shen}},
  \bibinfo{author}{\bibfnamefont{D.}~\bibnamefont{Andreucci}},
  \bibinfo{author}{\bibfnamefont{H.}~\bibnamefont{Hamm}}, \bibnamefont{and}
  \bibinfo{author}{\bibfnamefont{E.}~\bibnamefont{DiBenedetto}},
  \bibinfo{journal}{Biophys. J.} \textbf{\bibinfo{volume}{91}},
  \bibinfo{pages}{1192} (\bibinfo{year}{2006}).

\bibitem[{\citenamefont{Pugh~Jr and Lamb}(1992)}]{PughLamb1992}
\bibinfo{author}{\bibfnamefont{E.}~\bibnamefont{Pugh~Jr}} \bibnamefont{and}
  \bibinfo{author}{\bibfnamefont{T.}~\bibnamefont{Lamb}}, \bibinfo{journal}{J.
  Physiol.} \textbf{\bibinfo{volume}{449}}, \bibinfo{pages}{719}
  (\bibinfo{year}{1992}).

\bibitem[{\citenamefont{Rieke and Baylor}(1996)}]{RiekeBaylor1996}
\bibinfo{author}{\bibfnamefont{F.}~\bibnamefont{Rieke}} \bibnamefont{and}
  \bibinfo{author}{\bibfnamefont{D.}~\bibnamefont{Baylor}},
  \bibinfo{journal}{Biophys. J.} \textbf{\bibinfo{volume}{71}},
  \bibinfo{pages}{2553} (\bibinfo{year}{1996}).

\bibitem[{\citenamefont{Felber et~al.}(1996)\citenamefont{Felber, Breuer,
  Petruccione, Honerkamp, and Hofmann}}]{Felber1996}
\bibinfo{author}{\bibfnamefont{S.}~\bibnamefont{Felber}},
  \bibinfo{author}{\bibfnamefont{H.}~\bibnamefont{Breuer}},
  \bibinfo{author}{\bibfnamefont{F.}~\bibnamefont{Petruccione}},
  \bibinfo{author}{\bibfnamefont{J.}~\bibnamefont{Honerkamp}},
  \bibnamefont{and} \bibinfo{author}{\bibfnamefont{K.}~\bibnamefont{Hofmann}},
  \bibinfo{journal}{Biophys. J.} \textbf{\bibinfo{volume}{71}},
  \bibinfo{pages}{3051} (\bibinfo{year}{1996}).

\bibitem[{\citenamefont{Leskov et~al.}(2000)\citenamefont{Leskov, Klenchin,
  Handy, Whitlock, Govardovskii, Bownds, Lamb, Pugh~Jr, and
  Arshavsky}}]{Leskovetal2000}
\bibinfo{author}{\bibfnamefont{I.}~\bibnamefont{Leskov}},
  \bibinfo{author}{\bibfnamefont{V.}~\bibnamefont{Klenchin}},
  \bibinfo{author}{\bibfnamefont{J.}~\bibnamefont{Handy}},
  \bibinfo{author}{\bibfnamefont{G.}~\bibnamefont{Whitlock}},
  \bibinfo{author}{\bibfnamefont{V.}~\bibnamefont{Govardovskii}},
  \bibinfo{author}{\bibfnamefont{M.}~\bibnamefont{Bownds}},
  \bibinfo{author}{\bibfnamefont{T.}~\bibnamefont{Lamb}},
  \bibinfo{author}{\bibfnamefont{E.}~\bibnamefont{Pugh~Jr}}, \bibnamefont{and}
  \bibinfo{author}{\bibfnamefont{V.}~\bibnamefont{Arshavsky}},
  \bibinfo{journal}{Neuron} \textbf{\bibinfo{volume}{27}}, \bibinfo{pages}{525}
  (\bibinfo{year}{2000}).

\bibitem[{\citenamefont{Hamer et~al.}(2003)\citenamefont{Hamer, Nicholas,
  Tranchina, Liebman, and Lamb}}]{Hamer2003}
\bibinfo{author}{\bibfnamefont{R.}~\bibnamefont{Hamer}},
  \bibinfo{author}{\bibfnamefont{S.}~\bibnamefont{Nicholas}},
  \bibinfo{author}{\bibfnamefont{D.}~\bibnamefont{Tranchina}},
  \bibinfo{author}{\bibfnamefont{P.}~\bibnamefont{Liebman}}, \bibnamefont{and}
  \bibinfo{author}{\bibfnamefont{T.}~\bibnamefont{Lamb}},
  \bibinfo{journal}{J.~Gen.~Physiol.} \textbf{\bibinfo{volume}{122}},
  \bibinfo{pages}{419} (\bibinfo{year}{2003}).

\bibitem[{\citenamefont{Reingruber and Holcman}(2008)}]{ReingruberHolcman2007}
\bibinfo{author}{\bibfnamefont{J.}~\bibnamefont{Reingruber}} \bibnamefont{and}
  \bibinfo{author}{\bibfnamefont{D.}~\bibnamefont{Holcman}},
  \bibinfo{journal}{Biophys. J.} \textbf{\bibinfo{volume}{94}},
  \bibinfo{pages}{1954} (\bibinfo{year}{2008}).

\bibitem[{\citenamefont{Holcman and Korenbrot}(2005)}]{HolcmanKorenbrot2005}
\bibinfo{author}{\bibfnamefont{D.}~\bibnamefont{Holcman}} \bibnamefont{and}
  \bibinfo{author}{\bibfnamefont{J.}~\bibnamefont{Korenbrot}},
  \bibinfo{journal}{J. Gen. Physiol.} \textbf{\bibinfo{volume}{125}},
  \bibinfo{pages}{641} (\bibinfo{year}{2005}).

\bibitem[{\citenamefont{Hamer et~al.}(2005)\citenamefont{Hamer, Nicholas,
  Tranchina, Lamb, and Jarvinen}}]{Hamer2005}
\bibinfo{author}{\bibfnamefont{R.}~\bibnamefont{Hamer}},
  \bibinfo{author}{\bibfnamefont{S.}~\bibnamefont{Nicholas}},
  \bibinfo{author}{\bibfnamefont{D.}~\bibnamefont{Tranchina}},
  \bibinfo{author}{\bibfnamefont{T.}~\bibnamefont{Lamb}}, \bibnamefont{and}
  \bibinfo{author}{\bibfnamefont{J.}~\bibnamefont{Jarvinen}},
  \bibinfo{journal}{Vis. Neurosci.} \textbf{\bibinfo{volume}{22}},
  \bibinfo{pages}{417} (\bibinfo{year}{2005}).

\bibitem[{\citenamefont{Caruso et~al.}(2005)\citenamefont{Caruso, Khanal,
  Alexiadis, Rieke, Hamm, and DiBenedetto}}]{DiBenedetto2005}
\bibinfo{author}{\bibfnamefont{G.}~\bibnamefont{Caruso}},
  \bibinfo{author}{\bibfnamefont{H.}~\bibnamefont{Khanal}},
  \bibinfo{author}{\bibfnamefont{V.}~\bibnamefont{Alexiadis}},
  \bibinfo{author}{\bibfnamefont{F.}~\bibnamefont{Rieke}},
  \bibinfo{author}{\bibfnamefont{H.}~\bibnamefont{Hamm}}, \bibnamefont{and}
  \bibinfo{author}{\bibfnamefont{E.}~\bibnamefont{DiBenedetto}},
  \bibinfo{journal}{IEE proc.-Syst. Biol.} \textbf{\bibinfo{volume}{153}},
  \bibinfo{pages}{119} (\bibinfo{year}{2005}).

\bibitem[{\citenamefont{Fain et~al.}(2001)\citenamefont{Fain, Matthews,
  Cornwall, and Koutalos}}]{RevFainMatthews2001}
\bibinfo{author}{\bibfnamefont{G.}~\bibnamefont{Fain}},
  \bibinfo{author}{\bibfnamefont{H.}~\bibnamefont{Matthews}},
  \bibinfo{author}{\bibfnamefont{M.}~\bibnamefont{Cornwall}}, \bibnamefont{and}
  \bibinfo{author}{\bibfnamefont{Y.}~\bibnamefont{Koutalos}},
  \bibinfo{journal}{Physiological Reviews} \textbf{\bibinfo{volume}{81}}
  (\bibinfo{year}{2001}).

\bibitem[{\citenamefont{Koutalos et~al.}(1995)\citenamefont{Koutalos, Nakatani,
  and Yau}}]{Koutalosetal1995}
\bibinfo{author}{\bibfnamefont{Y.}~\bibnamefont{Koutalos}},
  \bibinfo{author}{\bibfnamefont{K.}~\bibnamefont{Nakatani}}, \bibnamefont{and}
  \bibinfo{author}{\bibfnamefont{K.-W.} \bibnamefont{Yau}},
  \bibinfo{journal}{Biophys. J.} \textbf{\bibinfo{volume}{68}},
  \bibinfo{pages}{373} (\bibinfo{year}{1995}).

\bibitem[{\citenamefont{Olson and Pugh~Jr}(1993)}]{OlsonPugh1993}
\bibinfo{author}{\bibfnamefont{A.}~\bibnamefont{Olson}} \bibnamefont{and}
  \bibinfo{author}{\bibfnamefont{E.}~\bibnamefont{Pugh~Jr}},
  \bibinfo{journal}{Biophys. J.} \textbf{\bibinfo{volume}{65}},
  \bibinfo{pages}{1335} (\bibinfo{year}{1993}).

\bibitem[{\citenamefont{Nickell et~al.}(2007)\citenamefont{Nickell, Park,
  Baumeister, and Palczewski}}]{Nickeletal2007}
\bibinfo{author}{\bibfnamefont{S.}~\bibnamefont{Nickell}},
  \bibinfo{author}{\bibfnamefont{P.~S.-H.} \bibnamefont{Park}},
  \bibinfo{author}{\bibfnamefont{W.}~\bibnamefont{Baumeister}},
  \bibnamefont{and}
  \bibinfo{author}{\bibfnamefont{K.}~\bibnamefont{Palczewski}},
  \bibinfo{journal}{J. of Cell Biol.} \textbf{\bibinfo{volume}{177}},
  \bibinfo{pages}{917} (\bibinfo{year}{2007}).

\bibitem[{\citenamefont{Naqvi}(1974)}]{Naqvi1974}
\bibinfo{author}{\bibfnamefont{K.~R.} \bibnamefont{Naqvi}},
  \bibinfo{journal}{Chemical Physics Letters} \textbf{\bibinfo{volume}{28}},
  \bibinfo{pages}{280} (\bibinfo{year}{1974}).

\bibitem[{\citenamefont{Torney and McConnell}(1983)}]{TorneyMcConnell1983}
\bibinfo{author}{\bibfnamefont{D.~C.} \bibnamefont{Torney}} \bibnamefont{and}
  \bibinfo{author}{\bibfnamefont{H.~M.} \bibnamefont{McConnell}},
  \bibinfo{journal}{Proc. of the Royal Society A}
  \textbf{\bibinfo{volume}{387}}, \bibinfo{pages}{147} (\bibinfo{year}{1983}).

\bibitem[{\citenamefont{Singer et~al.}(2006{\natexlab{a}})\citenamefont{Singer,
  Schuss, Holcman, and Eisenberg}}]{HolcmanetalNE1}
\bibinfo{author}{\bibfnamefont{A.}~\bibnamefont{Singer}},
  \bibinfo{author}{\bibfnamefont{Z.}~\bibnamefont{Schuss}},
  \bibinfo{author}{\bibfnamefont{D.}~\bibnamefont{Holcman}}, \bibnamefont{and}
  \bibinfo{author}{\bibfnamefont{B.}~\bibnamefont{Eisenberg}},
  \bibinfo{journal}{J. Stat. Phys.} \textbf{\bibinfo{volume}{122}},
  \bibinfo{pages}{437} (\bibinfo{year}{2006}{\natexlab{a}}).

\bibitem[{\citenamefont{Singer et~al.}(2006{\natexlab{b}})\citenamefont{Singer,
  Schuss, Holcman, and Eisenberg}}]{HolcmanetalNE2}
\bibinfo{author}{\bibfnamefont{A.}~\bibnamefont{Singer}},
  \bibinfo{author}{\bibfnamefont{Z.}~\bibnamefont{Schuss}},
  \bibinfo{author}{\bibfnamefont{D.}~\bibnamefont{Holcman}}, \bibnamefont{and}
  \bibinfo{author}{\bibfnamefont{B.}~\bibnamefont{Eisenberg}},
  \bibinfo{journal}{J. Stat. Phys.} \textbf{\bibinfo{volume}{122}},
  \bibinfo{pages}{465} (\bibinfo{year}{2006}{\natexlab{b}}).

\bibitem[{\citenamefont{Singer et~al.}(2006{\natexlab{c}})\citenamefont{Singer,
  Schuss, and Holcman}}]{HolcmanetalNE3}
\bibinfo{author}{\bibfnamefont{A.}~\bibnamefont{Singer}},
  \bibinfo{author}{\bibfnamefont{Z.}~\bibnamefont{Schuss}}, \bibnamefont{and}
  \bibinfo{author}{\bibfnamefont{D.}~\bibnamefont{Holcman}},
  \bibinfo{journal}{J. Stat. Phys.} \textbf{\bibinfo{volume}{122}},
  \bibinfo{pages}{491} (\bibinfo{year}{2006}{\natexlab{c}}).

\bibitem[{\citenamefont{Holcman and Schuss}(2008)}]{HolcmanSchuss2008_cluster}
\bibinfo{author}{\bibfnamefont{D.}~\bibnamefont{Holcman}} \bibnamefont{and}
  \bibinfo{author}{\bibfnamefont{Z.}~\bibnamefont{Schuss}},
  \bibinfo{journal}{J. Phys. A: Math. Theor.} \textbf{\bibinfo{volume}{41}},
  \bibinfo{pages}{155001} (\bibinfo{year}{2008}).

\bibitem[{\citenamefont{Holcman and Korenbrot}(2004)}]{HolcmanKorenbrot2004}
\bibinfo{author}{\bibfnamefont{D.}~\bibnamefont{Holcman}} \bibnamefont{and}
  \bibinfo{author}{\bibfnamefont{J.}~\bibnamefont{Korenbrot}},
  \bibinfo{journal}{Biophys. J.} \textbf{\bibinfo{volume}{86}},
  \bibinfo{pages}{2566} (\bibinfo{year}{2004}).

\bibitem[{\citenamefont{Dumke et~al.}(1994)\citenamefont{Dumke, Arshavsky,
  Calvert, Bownds, and PUGH~Jr}}]{Dumkeetal1995}
\bibinfo{author}{\bibfnamefont{C.}~\bibnamefont{Dumke}},
  \bibinfo{author}{\bibfnamefont{V.}~\bibnamefont{Arshavsky}},
  \bibinfo{author}{\bibfnamefont{P.}~\bibnamefont{Calvert}},
  \bibinfo{author}{\bibfnamefont{M.}~\bibnamefont{Bownds}}, \bibnamefont{and}
  \bibinfo{author}{\bibfnamefont{E.}~\bibnamefont{PUGH~Jr}},
  \bibinfo{journal}{J. Gen. Physiol.} \textbf{\bibinfo{volume}{103}},
  \bibinfo{pages}{1071} (\bibinfo{year}{1994}).

\bibitem[{\citenamefont{Gardiner}(2003)}]{BookGardiner}
\bibinfo{author}{\bibfnamefont{C.}~\bibnamefont{Gardiner}},
  \emph{\bibinfo{title}{Handbook of Stochastic Methods}}
  (\bibinfo{publisher}{Springer}, \bibinfo{year}{2003}), \bibinfo{edition}{3rd}
  ed.

\end{thebibliography}

\end{document}